\documentclass[3p,singlecolumn]{elsarticle}

\usepackage{geometry}
\usepackage{multirow}
\usepackage{hyperref}
\usepackage{mathtools}
\usepackage[utf8]{inputenc}
\usepackage[T1]{fontenc}
\usepackage{mathtools}
\usepackage[thinc]{esdiff}
\usepackage{comment}
\DeclareUnicodeCharacter{2113}{\ensuremath{\ell}}

\usepackage{graphicx}
\usepackage{amsmath,amssymb,amsfonts}

\usepackage{subcaption}
\usepackage{amsthm}
\usepackage{multirow}

\usepackage{mathtools}
\usepackage{systeme}

\usepackage{lineno}
\usepackage{xcolor}

\DeclareMathOperator{\rank}{rank}

\journal{Engineering Applications of Artificial Intelligence}

\begin{document}

\begin{frontmatter}

\title{CARONTE: a Physics-Informed Extreme Learning Machine-Based Algorithm for Plasma Boundary Reconstruction in Magnetically Confined Fusion Devices}



\author[unina,create]{Federico~Fiorenza}\ead{federico.fiorenza@unina.it}    
\author[unina,create]{Sara~Dubbioso}\ead{sara.dubbioso@unina.it}    
\author[unina,create]{Gianmaria~De~Tommasi}\ead{detommas@unina.it}    
\author[unina,create]{Alfredo~Pironti}\ead{pironti@unina.it}    

\address[unina]{Dipartimento di Ingegneria Elettrica e delle Tecnologie dell'Informazione, Universit\`a degli Studi di Napoli Federico II, Napoli, Italy}
\address[create]{Consorzio CREATE, Napoli, Italy}

\begin{abstract}
In this work, we propose a novel physics-informed neural network-based algorithm for real-time plasma boundary reconstruction in tokamak devices. The approach is based on a single Extreme Learning Machine network used to solve the homogeneous Grad–Shafranov equation, which is required to identify the plasma boundary. This architecture enables the real-time training of the network parameters using the available magnetic sensor data and, consequently, dynamically adapting the network output to the evolving plasma equilibrium. We demonstrate that,
the network performs accurate plasma boundary reconstruction for complex configurations, outperforming well-established methods, such as the~algorithm used for decades at the~Joint European Torus, the world's largest tokamak, until it ceased operation in~2023. Indeed, compared to the latter, the proposed solution better generalizes the poloidal flux function, without requiring algorithm retuning across different plasma equilibria. The proposed neural network reconstructor demonstrates also greater robustness with respect to noise on the magnetic measurements. Moreover, this method takes advantage of the generalization power of neural networks but without the need for extensive, time-consuming training based on a huge amount of experimental data, making its implementation on existing devices straightforward.

\end{abstract}

\end{frontmatter}


\section{Introduction}\label{section:Introduction}
\begin{figure}[!thb]
    \centering
    \includegraphics[width=.5\linewidth]{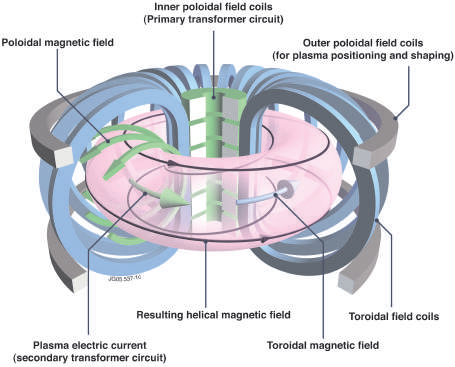}
    \caption{Simplified scheme of a tokamak fusion device.}
    \label{fig:Tokamak}
\end{figure}

In the continuous quest for energy sources, nuclear fusion is foreseen as an alternative that could be available in the next century. A big effort in this direction started soon after World War II, leading to big International projects that aim at demonstrating the feasibility of a nuclear fusion power plant in the next future. Among the various International efforts, the ITER project involving~EU, India, People's Republic of China, Russia, South Korea and USA, and which is currently under construction in France, is the most ambitious one, being designed to be the first magnetic confinement device to produce a net surplus of fusion energy~\cite{ITER}. While waiting the completion of~ITER construction, other satellite projects have been setup, such as the joint EU-Japan project~JT-60SA, currently the World's largest operating fusion device, and the Divertor Tokamak Test~(DTT) facility in Italy, that will test the strategy to handle the heat exhaust in future reactors~\cite{romanelli2024divertor}. All these projects exploit \emph{magnetic confined fusion} to reach the target. Indeed, nuclear fusion occurs when nuclei of light elements overcome their mutual electrostatic repulsion, known as the Coulomb barrier, and collide with sufficient energy to combine into a heavier nucleus. This process releases a substantial amount of energy, primarily in the form of kinetic energy. The probability of the fusion reaction to occur under specific conditions is represented by the so-called cross-section. The high cross-section at relatively low temperatures of the mixture of deuterium~(D) and tritium~(T) justifies its wide preference. 
To achieve fusion conditions, the~D-T gas mixture needs to be heated to extremely high temperatures, i.e., $T\geq10^8~K$, at which it becomes a fully ionized gas called \emph{plasma}. The plasma must be maintained and confined for a time sufficient to provide the kinetic energy necessary for the nuclei to overcome the Coulomb forces. 
Tokamaks are pulsed fusion devices~\cite{wesson2011tokamaks} and exploit such a principle by providing confinement by means of strong toroidal and poloidal magnetic fields, which are generated by the corresponding set of external coils shown in the simplified scheme of Fig.~\ref{fig:Tokamak}. These two magnetic components bend the field lines, resulting in a helical magnetic field that has the main advantage of avoiding particle end-losses. The Poloidal Field~(PF) is mainly given by the current flowing in the plasma itself, whereas the poloidal magnets provide the necessary magnetic field to control the so-called \emph{plasma equilibrium}. Indeed, they induce the plasma current and regulate its position and shape during each phase of the discharge (ramp-up, flattop and ramp-down), aiming at avoiding disruptions and optimizing the performance~\cite{AriolaPironti:Springer}.



Such a plasma magnetic control task is further complicated by the fact that, to enhance performance in tokamak configurations, elongated shapes are usually pursued, which exhibit vertical instability~\cite{walker2009feedback}. As a consequence, plasma equilibrium control calls for integrated control approaches, as it is an intrinsically~MIMO problem where multiple actuators are shared to fulfill multiple objectives. 
In this context, several solutions have been proposed and tested on experimental devices in the last decades. In early devices running \emph{limiter} configurations~\cite[Tutorial~7]{beghi2005advances} with circular plasmas, the problem was decoupled by dedicating different sets of~PF coils to independently control plasma current and position by means of~SISO loops~\cite[Ch.~6]{AriolaPironti:Springer}. Such an approach cannot be pursued in modern devices aiming at controlling high-performance elongated \emph{diverted} plasmas. Therefore~MIMO architectures have been designed following model-based approach~\cite{de2024control}, exploiting a wide range of techniques, including robust control~$\mathcal{H}_{\infty}$~\cite{ariola2002design}, geometric approach based on singular value decomposition~\cite{ambrosino2008design}, adaptive control~\cite{inoue2021development}, up to data-driven approaches, such as reinforcement learning~\cite{degrave2022magnetic,dubbioso2023deep}. 

Despite the different adopted strategies, every equilibrium control algorithm relies on the real-time reconstruction of the plasma boundary to feedback the controlled shape descriptors, which can be either poloidal-flux difference at given control points~\cite{yuan2013plasma} (also called \emph{isoflux} control strategy) or plasma-wall gap along given segments~\cite{ambrosino2009design}. 

A possible approach to retrieve these quantities in real-time implies the \emph{online} solution of the~2D magnetohydrodynamics equilibrium~(MHD,~\cite[Ch.~11]{freidberg2008plasma}) during the plasma discharge, by using one of the many available plasma equilibrium codes, such as RT-EFIT~\cite{ferron1997real}, LIUQE~\cite{moret2015tokamak} and PEFIT~\cite{huang2018improvement}. However, such codes are mainly used for offline analysis since they can guarantee good accuracy when run in iterative mode. 
When used in real-time, they are run in \emph{single-iteration} mode, to meet the required sampling time, which typically ranges between~500~$\mu s$ and~$1$~ms.

Methods based on a simplified representation of the plasma by means of parametrized current distributions~\cite{kurihara2000new,fiorenza2026preliminary}, or on local approximations of the poloidal flux~\cite{sartori2003jet} permit not only to meet the real-time requirement, but also to improve the quality of the reconstruction, 
if compared with the one returned by a \emph{single-iteration} of equilibrium codes.

\medskip

In this work, we present the CARONTE~(Create Ai RecONsTructor of plasma surfacE) algorithm for real-time plasma boundary reconstruction, which is based on a single Physics-Informed Neural Network~(PINN). In particular, one physics-informed Extreme Learning Machine (ELM) Neural Network (NN) is adopted, which belongs to the family of fast-learning Reservoir Computing Networks~\cite{lukovsevivcius2009reservoir}, and projects the inputs in a random and non-linear fashion onto a high-dimensional feature space. Indeed, a unique hidden layer of nonlinear neurons, known as \emph{reservoir}, is randomly connected to the input nodes. The remaining connections between the hidden layer and the output nodes are updated during the training process. Hence, differently from traditional deep neural networks, ELM NNs do not require iterative training,  also resulting in a faster learning time. ELM NNs have proved to be effective in many applications~\cite{ding2014extreme,dubbioso2024model}. In this paper, we exploit such a mechanism to propose an algorithm for plasma boundary reconstruction based on real-time training of the network hyper-parameters by using the available magnetic measurements. This mechanism leads to a plasma boundary reconstructor that is dynamically adapted to evolving plasma equilibrium. 
As shown in the following sections, the proposed approach proved to be effective in reconstructing the plasma boundary in both in small-size and almost circularly shaped machines, such as RFX-mod2~\cite{marrelli2019upgrades}, as well as in large-size tokamaks such as~DTT, where advanced plasma configurations with \emph{exotic} shapes are foreseen~\cite{ambrosino2019magnetic}. For this latter case study, the CARONTE algorithm outperforms a~XLOC-like approach, being XLOC the algorithm used for decades at the~Joint European Torus (JET) tokamak~\cite{sartori2003jet}. It is worth remarking that the performance evaluation has been carried out considering the equilibria of the CREATE codes~\cite{albanese1998linearized,albanese2015create} as ground truth, since these codes have been successfully validated on many tokamak devices, including JET~\cite{ambrosino2008design}, TCV~\cite{frattolillo2025implementation}, EAST~\cite{albanese2017iter}, JT-60SA~\cite{fiorenza2026preliminary} and ITER~\cite{ambrosino2009design}, other than~RFX-mod2 and~DTT.

The use of machine learning to solve plasma equilibrium and boundary reconstruction has already been considered in the literature. However, all the proposed approaches are rather \emph{standard} ones, based on offline training of deep~NNs, which are typically demanding both in terms of computational time and amount of training data that are needed to achieve satisfactory results. In both~\cite{wang2024neural,kaltsas2022neural} a single physics-informed deep neural network is trained on numerical data, while in~\cite{joung2023gs} an architecture with two interacting networks is considered, one for the Maxwell equation and the other for the force balance included in~MHD. Authors of~\cite{bonotto2024reconstruction} consider a single physics-informed~NN to solve the equilibrium, which is then classified as limiter or diverted by another network; the outputs of these two networks are then used by a third one to compute the plasma boundary. In~\cite{wan2023machine} deep-NN are used to estimate the plasma boundary, rather than the whole equilibrium, but also in this case the approach requires a not negligible effort to train the network. Moreover, a fully data-driven approach based on Bayesian inference, Gaussian Process Regression, and Markov Chain Monte Carlo has also been proposed in~\cite{nishizawa2024equilibrium}. However, as stated by the authors themselves, such an approach is not suitable for real-time applications, given the required~\emph{``$\ldots$significant
computation expense$\ldots$"}. 

Summarizing, if compared to all the above approaches, being based on a single physics-informed~ELM network, our approach presents the following advantages, which make it suitable for real-time implementation:
\begin{enumerate}

\item \textbf{Online training}; unlike deep~NN, no offline training on large datasets is required to train the single ELM network, which is trained online at each sampling time, by a single matrix multiplication, on the basis of the latest magnetic field and flux measurements.
\item \textbf{Minimal computational resources}; since the only parameters to be trained are the weights of the output layer, a single~ELM has orders-of-magnitude fewer parameters than deep architectures; hence, for the considered problem, it is possible to meet the required timescale for real-time implementation.
\item \textbf{Implementation Simplicity}; being based on a single~NN, no coupling between multiple networks or staged pipelines are present in the proposed architecture.
\end{enumerate}

For the sake of completeness, it is important to remark that the proposed approach focuses solely on plasma boundary reconstruction, although further research is ongoing to extend it to real-time reconstruction of the whole plasma equilibrum.

\medskip

The paper is structured as follows: the next section introduces some background about plasma magnetic diagnostics. In particular, the Grad-Shafranov equation is presented, whose solutions are the plasma equilibria under the assumption of axis-symmetry, which allows to recast a~3D problem into a~2D one. The same section briefly describes the various magnetic sensors that are typically installed in tokamaks, as well as the relevant plasma quantities that need to be reconstructed and controlled in real-time. Section~\ref{section:BoundaryReconstruction} describes the two main model-based approaches for plasma boundary reconstruction, namely the filamentary approach and a~XLOC-like algorithm, since these are considered for the comparison with the~CARONTE algorithm, which is described in details in Section~\ref{section:PINNReconstruction}, and whose validation is presented in Section~\ref{section:validation}. A conclusive discussion is given in Section~\ref{section:Conclusions}.

\section{Plasma equilibrium and magnetic diagnostics}\label{section:preliminaries}
This section first introduces the Grad-Shafranov equation, whose solution is the poloidal flux function~$\psi$ that fully describes~2D~MHD plasma equilibria. A short discussion about the so-called plasma magnetic diagnostics, i.e., of the algorithms that are used to estimate the plasma quantities relevant for magnetic control starting from the available measurements, is also given in this section.

\subsection{The Grad-Shafranov equation}\label{subsection:EquilibriumGS}

\begin{figure}[t]
	 \centering
	 \includegraphics[width=0.3\linewidth]{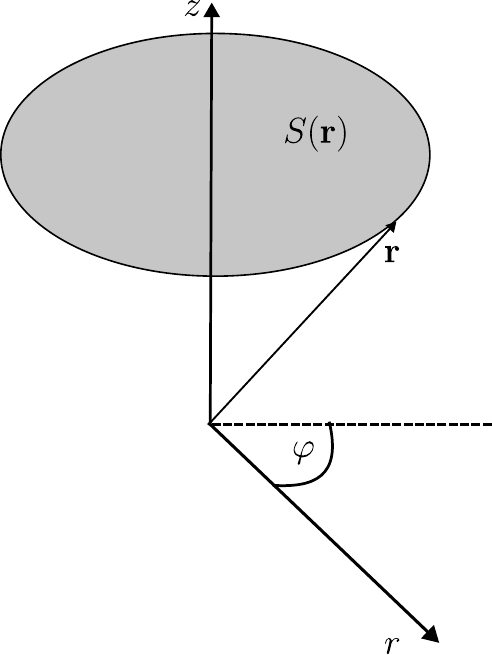}
	 \caption{The cylindrical coordinate system and the surface $S(\mathbf{r})$ obtained by rotating $\mathbf{r}$ around the $r=0$ axis.}
	 \label{fig:CilinderCoordinates}
\end{figure}

An important physical quantity to describe the plasma equilibrium in a tokamak device is the poloidal magnetic flux. Its behaviour is described by the Grad-Shafranov~(GS) equation.   
To introduce the poloidal flux function, let $(r,\varphi,z)$ be a three-dimensional cylindrical coordinate system, as shown in Fig.~\ref{fig:CilinderCoordinates}. 
Since a tokamak is an axisymmetric toroidal machine, it is assumed for the generic vector $\mathcal{A}$ 
\begin{equation}\label{eq:AxisSymm}
    \frac{\partial}{\partial \varphi}\mathcal{A} = 0\,,
\end{equation}
reducing the plasma equilibrium to a two-dimensional problem.

Let us now define the poloidal flux function. It is defined as 
\begin{equation}\label{eq:PolFlux}
    \psi(\mathbf{r}) = \frac{1}{2\pi}\int_{S(\mathbf{r})}\mathbf{B}\cdot\mathrm{d}\mathbf{S}\,,
\end{equation}

where $S(\mathbf{r})$ is a surface such that its contour $\partial S$ is a circumference obtained by rotating the point $\mathbf{r}$ around the~$r=0$ axis~(see Fig.~\ref{fig:CilinderCoordinates}). Hence, it is $\psi(\mathbf{r})=\psi(r,z)$.
By choosing $S(\mathbf{r})$ perpendicular to the  $r=0$ axis, it is possible to write the surface integral in Eq.~\eqref{eq:PolFlux} as 
\begin{equation}\label{eq:PolFlux2}
    \psi(\mathbf{r}) = \frac{1}{2\pi}\int_{0}^r\int_{0}^{2\pi}B_z(\rho,z)\rho\mathrm{d}\rho\mathrm{d}z = \int_0^r\rho B_z(\rho,z)d\rho\,,
\end{equation}
where $\mathbf{B}=B_r\hat{\mathbf{r}} + B_z\hat{\mathbf{z}} + B_\varphi\hat{\boldsymbol{{\varphi}}} =\mathbf{B_p}+ B_\varphi\hat{\boldsymbol{{\varphi}}}$, where $\mathbf{B_p}$ is the poloidal magnetic field.

When~\ref{eq:AxisSymm} holds, then Gauss's Law for the magnetic field in cylindrical coordinates becomes 
\begin{equation}\label{eq:GaussCylindric}
    \frac{1}{r}\frac{\partial}{\partial r}rB_r + \frac{\partial}{\partial z}B_z=0\,.
\end{equation}
By differentiating Eq.~\eqref{eq:PolFlux2} with respect to $r$ and $z$, and by taking into account Eq.~\eqref{eq:GaussCylindric}, the following relations between the poloidal flux function and the magnetic field can be derived: 
\begin{equation}\label{eq:FluxFieldRel}
\begin{cases}
    \frac{\partial}{\partial r}\psi=rB_z\\
    \frac{\partial}{\partial z}\psi=-rB_r
    \end{cases}\,.
\end{equation}
Moreover, from Eq.~\eqref{eq:FluxFieldRel} it also follows that
\begin{equation*}
    \mathbf{B}\cdot\nabla\psi =0\,.
\end{equation*}
Hence, the magnetic field lines lie on constant poloidal flux surfaces. 
Let us define the $\Delta^*$ operator as:
\begin{equation}\label{eq:NablaStar}
    \Delta^*f = r\frac{\partial}{\partial r}\left(\frac{1}{r}\frac{\partial}{\partial r}f\right)+\frac{\partial^2}{\partial z^2}f\,.
\end{equation}
Given a current source, the GS equation is
\begin{equation}\label{eq:GDeq}
    \Delta^*\psi=-\mu_0rJ_\varphi\,,
\end{equation}
where $J_\varphi$ is the toroidal current density whose expression is different depending on the region of the poloidal plane of interest. Clearly, in the vacuum, the poloidal flux function satisfies the homogeneous GS equation
\begin{equation}\label{eq:GDeqHomo}
    \Delta^*\psi=0\,.
\end{equation}
In the active coils region, $J_\varphi$ is determined by the toroidal current flowing through them. Finally, in the plasma region, the toroidal current depends on the pressure profile and the toroidal component of the magnetic induction field. 

The GS equation reads as follows in the different domains:
\begin{equation}\label{eq:GDsys}
    \begin{cases}
        \Delta^*\psi=0 & \text{in the vacuum region}\\
        \Delta^*\psi=-\mu_0rJ_{ext} & \text{in the active coils}\\
        \Delta^*\psi=-\mu_0rJ_p\ & \text{in the plasma region}\\
        \psi(0,z) = 0 \\
        \displaystyle\lim_{r \to \infty} \psi(r,z) = 0
    \end{cases}
\end{equation}
where $J_p$ is the plasma current density, and $J_{ext}$ is the toroidal current density in the active coils. Usually, assuming an analytical description of $J_p$, numerical codes can solve these non-linear partial differential equations where the boundary of the plasma region is part of the unknown to be determined. 

\subsection{Magnetic diagnostics}\label{subsection:magDiag}
All the plasma quantities regulated by the magnetic control system, i.e., the plasma boundary descriptors as well as the plasma current and centroid position, cannot be directly measured, and they are estimated starting from the available measurements. Typically, magnetic diagnostics system refers to the set of magnetic sensors and algorithms needed to estimate the plasma quantities of interest. The main plasma parameters reconstructed in real-time during a discharge, as well as the sensors needed, are presented in what follows.
\subsubsection{Plasma current}
The induced plasma current is one of the main parameters that defines the whole plasma equilibrium. It is the main source of the poloidal magnetic field, and its value is crucial for particle confinement. The total plasma current is the current flowing through the poloidal plane in the plasma region, and it is defined as follows:
\begin{equation*}
    I_p = \int_{\Omega_p}J_\varphi \mathrm{d}S\,,
\end{equation*}
where $\Omega_p$ is the plasma region (the plasma region is the one contained inside the plasma boundary). The plasma current can be either reconstructed via the magnetic field measurements or directly measured with dedicated probes, called Rogowski coils. 
\subsubsection{Plasma current centroid position}
The plasma current centroid is the center of the plasma current distribution. In particular, 
\begin{equation*}
    z_c=\frac{1}{I_p}\int_{\Omega_p}zJ_\varphi\mathrm{d}S
\end{equation*}
 is used to characterize the plasma vertical position, while 
 \begin{equation*}
    r_c=\sqrt{\frac{1}{I_p}\int_{\Omega_p}r^2J_\varphi\mathrm{d}S}
\end{equation*}
characterizes the plasma radial position. 

Estimating and controlling the plasma current centroid position is critical for the proper operation of a tokamak. In fact, this quantity plays a key role in controlling the plasma's position within the chamber. It's also important to note that, during the initial phase of a tokamak pulse, the plasma current centroid position is often controlled independently of shape control. Furthermore, for vertically unstable plasmas, controlling the centroid position $z_c$ is crucial for maintaining stability. 

\subsubsection{The plasma boundary}
The identification of the plasma boundary is essential for real-time shape control. In fact, plasma shape descriptors are needed for the control system to achieve the desired plasma configuration. The plasma boundary is identified with one of the constant level lines of the poloidal flux function~(i.e.,~the edge of magnetic surfaces), specifically with the Last Closed Flux Surface~(LCFS) entirely contained within the vacuum vessel. 
\begin{figure}
    \centering
    \includegraphics[width=.5\linewidth]{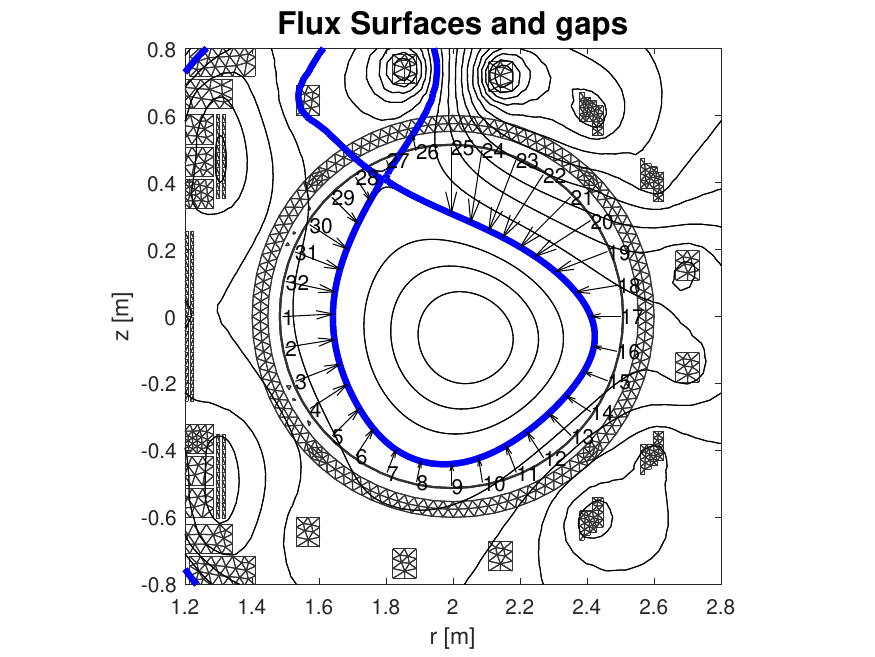}
    \caption{Plasma flux surfaces for a RFX-mod2 diverted plasma computed with CREATE-L. The LCFS highlighted in blue defines the plasma boundary. The arrows represent the plasma-wall gaps.}
    \label{fig:FluxSurf}
\end{figure}

The plasma can be either \emph{limiter} or \emph{diverted}. If the plasma boundary and the tokamak first wall (also called limiter) have a point in common, the plasma is called a limiter plasma. On the other hand, if the plasma boundary and the first wall do not share any point in common, the plasma is called a diverted plasma~(see also~Tutorial~7 in~\cite{beghi2005advances}). In the latter case, the boundary separates the closed and open flux surfaces. Moreover, the~LCFS is characterised by one or more saddle points (also called \emph{X-points} or null-points) for the poloidal flux function, therefore, in these points, the poloidal magnetic field is zero. For diverted plasmas, the plasma boundary continues after an X-point with two baffles. The intersections between the baffles and the first wall are called \emph{strike points}. Most of the power produced by the plasma is concentrated here; therefore, precise estimation and control of the strike points is fundamental to control the heat flux on the plasma-facing, components~\cite{ambrosino2008plasma}. As an example, Fig.~\ref{fig:FluxSurf} shows the isoflux surfaces on the poloidal plane of the RFX-mod2 device, highlighting the~LCFS for a diverted plasma.

To identify the plasma boundary, it is sufficient to solve the~GS equation in the vacuum, knowing the flux and the magnetic field at the measurement points~$(\mathbf{r_\psi},\mathbf{z_\psi})$ and~$(\mathbf{r_b},\mathbf{z_b})$:
\begin{equation}\label{eq:BoundRec}
    \begin{cases}
        \Delta^*\psi=0 \\
        \psi(\mathbf{r_\psi},\mathbf{z_\psi})=\mathbf{m_\psi}\\
        \mathbf{B_p}(\mathbf{r_b},\mathbf{z_b},\mathbf{\theta_B})=\mathbf{m_{B}}
    \end{cases}\,,
\end{equation}
where $\mathbf{m_\psi}$ and $\mathbf{m_{B}}$ are the flux and magnetic field measurements coming from the corresponding probes (see Section~\ref{subsection:sensors}).

Since the poloidal flux monotonically increases from the edge towards the center of the plasma, the value of the flux at the boundary $\psi^b$, to distinguish between limiter and diverted configurations, can be easily obtained by taking the maximum between the flux at the first wall and the flux at the X-points.

 Once the plasma edge is determined, the \emph{gaps} are computed, i.e.~the distances between the plasma boundary and the first wall along pre-defined segments (see the arrows in Fig.~\ref{fig:FluxSurf}). These quantities are controlled to the desired values to control the plasma shape, if a gap-based control strategy is adopted~\cite{ambrosino2008design}.

Let $\mathbf{p}_1$ be a point chosen on the first wall, while $\mathbf{p}_2$ a point inside the chamber. Each gap $g$ is determined by the intersection $\mathbf{p}_b$ of the plasma boundary with the segment connecting the points $\mathbf{p}_1$ and $\mathbf{p}_2$. Therefore, it is 
\[\mathbf{p}_{b}=\mathbf{p}_1+g\frac{\mathbf{p}_1-\mathbf{p}_2}{||\mathbf{p}_1-\mathbf{p}_2||}\,,\]
 and 
 \[\psi(\mathbf{p}_b)=\psi^b\,,\]
 where $\psi^b$ is the flux at the boundary.
The estimation of these shape descriptors can be made only after an estimation of the magnetic flux inside the chamber is given. 
Therefore, to control the plasma shape, either the gaps are controlled to the desired value or, given a number of control points, the flux difference \(\psi^b-\psi(\mathbf{p}^*)\) is controlled to zero so that the plasma boundary passes through the point $\mathbf{p^*}$. 
Boundary reconstruction codes, which only solve~Eq.~\eqref{eq:BoundRec}, can provide accurate flux maps only in the vacuum. These codes are simpler and faster, making them suitable for real-time control. 
On the other hand, full-equilibrium codes solve Eq.~\eqref{eq:GDsys} and are generally more accurate and reliable. 
However, as a free-boundary problem, these codes are iterative and consequently slower. Therefore, they require optimization for real-time deployment and sufficient computational resources.

\subsubsection{Magnetic sensors}\label{subsection:sensors}
Flux loops consist of an open coil whose voltage across the two terminals is measured by means of a differential amplifier. It is a single open
loop of wire, making a complete toroidal turn, and crossing the poloidal plane at a point~$\mathbf{r}$. According to Faraday's law, this voltage is proportional to the time derivative of the total magnetic flux linked with the coil.
\begin{equation}\label{eq:FluxLoopDer}
    V_\psi(t) = -2\pi\frac{\partial}{\partial t}\psi(\mathbf{r},t)\,.
\end{equation}
By integrating the measured voltage $V_\psi$, from Eq.~\eqref{eq:FluxLoopDer}
\begin{equation*}
    \psi(\mathbf{r},t)=-\frac{1}{2\pi}\int_{t_0}^tV_\psi(\tau)\mathrm{d}\tau + \psi(\mathbf{r},t_0)\,.
\end{equation*}
Therefore, flux loops, together with an integrator circuit, give the time evolution of the poloidal flux function at point $\mathbf{r}$.

Regarding the magnetic field measurement, a pick-up coil consists of multiple windings of wire of small radius located at a point $\mathbf{r}$. Since the section $A$ is very small, the induced voltage is proportional to the time derivative of the magnetic field passing through the solenoid:
\begin{equation}\label{eq:PickUpDer}
    V_\mathbf{B} = -NA\frac{\partial}{\partial t}\mathbf{B_\theta}(\mathbf{r},t)
\end{equation}
where $N$ is the number of windings, $\theta$ is the angle with respect to the radial axis at which the pick-up coil is mounted. Again, by integrating Eq.~\eqref{eq:PickUpDer}, the magnetic field along the direction given by $\theta$ is estimated. Therefore, it is:
\begin{equation}\label{eq:MagSens}    
m_\mathbf{B}(r_\mathbf{B},z_\mathbf{B},\theta_\mathbf{B}) = B_r(r_\mathbf{B},z_\mathbf{B})\cos(\theta_\mathbf{B}) + B_z(r_\mathbf{B},z_\mathbf{B})\sin(\theta_\mathbf{B})\,.
\end{equation}
\section{Conventional approaches for real-time plasma boundary reconstruction}\label{section:BoundaryReconstruction}
As already mentioned, the real-time full-equilibrium codes are computationally demanding and require resources that may not be available. This is why, for shape control purposes, non-iterative boundary reconstruction codes are typically used in real-time for plasma magnetic control~\cite{kurihara2000new,sartori2003jet,ambrosino2002line} since, other than being fast, they can also provide a first estimation of the flux-map to be used by more complex codes. In this section, two well-known and assessed boundary reconstruction methods are briefly described. In Section~\ref{section:validation}, they are used as benchmark for the new proposed method. The first one is a simple filamentary method, which is easily implementable and gives accurate reconstruction specially in smaller devices. The second one is inspired by~XLOC, which is the code that has proved to be effective on large tokamaks, indeed, it has been used at JET for several years~\cite{sartori2003jet}.
\subsection{Filamentary approach}\label{subsection:filamentary}
In this section, we will briefly describe a simplified filamentary reconstruction code. Let $n_m$ be the number of available measurements. The key idea is that these filaments approximate the plasma contribution to the magnetic field and flux. Indeed, a number of filaments~$n_f~\ll~n_m$ is placed inside the chamber on an ellipse around its center.  Then, neglecting the passive currents contribution,
\begin{equation}\label{eq:FluxRec}
    \mathbf{m} = 
    \begin{pmatrix}
        \mathbf{m_{B}} \\
        \mathbf{m_\psi}
    \end{pmatrix} = \mathbf{\Tilde{C}_{I_f}}\cdot\mathbf{I_{f}} +  \mathbf{\Tilde{C}_{I_a}}\cdot\mathbf{I_a}\,,
\end{equation}
where matrix $\mathbf{\Tilde{C}_{I_f}}$ ($\mathbf{\Tilde{C}_{I_a}}$) links the currents in the $n_f$ ($n_a$) filaments to vector $ \mathbf{m}$ and are derived from Green's functions for Eq.~\ref{eq:GDeqHomo} (for more details check the Appendix in~\cite{AriolaPironti:Springer}). Eq.~\eqref{eq:FluxRec} can be modified in 
\begin{equation}\label{eq:FluxRecSystem}
    \begin{pmatrix}
        \mathbf{m}\\
        \Bar{\mathbf{I}}_a\\
        \mathbf{0}
    \end{pmatrix} 
    = \mathbf{\Theta}^{-1}
    \begin{pmatrix}
         \mathbf{\Tilde{C}_{I_a}} & \mathbf{\Tilde{C}_{I_f}} \\
         \mathbf{I_{n_a}} & \mathbf{0}\\
         \mathbf{0} &  \mathbf{I_{n_a}}
    \end{pmatrix}
    \begin{pmatrix}
        \mathbf{I}_a \\
        \mathbf{I}_{f}
    \end{pmatrix}\,,
\end{equation}

where $\mathbf{\Theta}$ is a weighting matrix based on the measurement error standard deviations, $ \Bar{\mathbf{I}}_a$ are the measured currents, and $\mathbf{0}$ zero matrix of appropriate dimensions. From~Eq.~\eqref{eq:FluxRecSystem} $\mathbf{I}_{f}$ and $\mathbf{I}_{a}$ can be easily derived. Then, the flux map can be reconstructed by summing at all points of a predefined grid, all the sources' (represented by $\mathbf{I}_{f}$ and~$\mathbf{I}_{a}$) contributions to the poloidal flux. It is worth noticing that the position of the filaments can be changed during the execution of the algorithm to cope with the changes in the plasma position, however, to improve performance in larger tokamaks, the filaments distribution should be adapted to the plasma shape.

Moreover, from this method, one can compute the plasma current as $I_p = \sum_{i=1}^{n_f} I_{f_i}$ and the current centroid position as 
\begin{equation*}
    z_p = \frac{\sum_{i=1}^{n_f} z_{f_i}I_{f_i}}{\sum_{i=1}^{n_f} I_{f_i}}\,, \quad
    r_p = \sqrt{\frac{\sum_{i=1}^{n_f} r_{f_i}^2I_{f_i}}{\sum_{i=1}^{n_f} I_{f_i}}}\,.
\end{equation*}

Finally note that the eddy currents contribution could be taken into account by placing a sufficient number of filaments inside the conductive passive structures. A more advanced version of this algorithm is described in~\cite{kurihara2000new}. It has been used during the JT-60SA first operation in~2023~\cite{fiorenza2026preliminary,szepesi:IAEA2025}.

\subsection{XLOC-like approach}\label{subsection:XLOC}
The XLOC code~\cite{sartori2003jet} has been used at JET to compute the poloidal flux function in the vacuum from the magnetic measurement by satisfying the~GS equation in the vacuum region. Here, we briefly describe the approach adopted by the XLOC-like algorithm that has been implemented and used as a benchmark in Section~\ref{section:PINNReconstruction}.

The starting point is to choose a basis of functions that solve~Eq.~\eqref{eq:GDeqHomo}. Polynomial solutions to the homogeneous~GS equation are:
\begin{equation*}
\chi_n(r,z)=
\begin{cases}
    1 & n=1 \\
    \sum_{k=0}^{\lfloor n/2 \rfloor -1 } (-4)^{-k}\frac{(n-1)!}{2k!(k+1)!(n-2k-2)!}r^{2k+2}z^{n-2k-2} & n\geq2
    \end{cases}\,,
\end{equation*}

where $r$ and $z$ are the coordinates in the poloidal plane. Note that in~\cite{sartori2003jet} the chosen basis of function did not satisfy the GS, but the coefficients were constrained to do so. Let $\boldsymbol{\chi}= \begin{pmatrix}
    \chi_1 & \chi_2 & \cdots & \chi_{\bar{N}}\end{pmatrix}^T$ be the chosen vector of basis functions, then the poloidal flux function can be approximated by using a linear combination of the elements of $\boldsymbol{\chi}$:
\begin{equation}\label{eq:polyXLOC}
    \psi(r,z) = \boldsymbol{\chi}^T(r,z)\cdot\mathbf{\bar{a}}\,,
\end{equation}
where $\mathbf{\bar{a}}\in\mathbb{R}^{\bar{N}}$ is the vector of coefficients to be determined. Since one polynomial does not fit the whole flux distribution outside the plasma well enough, more polynomials are used, each one valid in a different region of the chamber. Note that the number of basis functions to be used may vary by region. For instance, if the chamber is divided into $K$ regions, the total number of coefficients to be computed is $N=\sum_{i=1}^{K}\bar{N}_i$. Then, by exploiting\linebreak Eq.~\eqref{eq:FluxFieldRel}--\eqref{eq:MagSens}, the magnetic sensors can also be used to determine the coefficients vector $\mathbf{a}$. Thus, Eq.~\eqref{eq:polyXLOC} can be reworked as
\begin{equation}\label{eq:sysXLOC}
    \mathbf{M}\cdot\mathbf{a}=\mathbf{m}\,,
\end{equation}
where $\mathbf{a}\in\mathbb{R}^N$, $\mathbf{m}\in\mathbb{R}^{n_m}$ is the vector of the measurements of the poloidal flux and the magnetic field, and~$\mathbf{M}\in \mathbb{R}^{n_m\times N}$ is the matrix of the basis functions and their derivatives computed offline in the points where the sensors are located. 

Eq.~\eqref{eq:sysXLOC} alone does not guarantee the continuity of the poloidal flux function at the boundary of the regions. This is why a point on each segment dividing the regions is selected. In this point, the constraint is added that the flux computed with two polynomials of adjacent regions is equal:
\begin{equation}\label{eq:HCXLOC}
    \boldsymbol{\chi}^T_i(r_p,z_p)\cdot \mathbf{\bar{a}}_i = \boldsymbol{\chi}^T_{i+1}(r_p,z_p)\cdot \mathbf{\bar{a}}_{i+1}\,.
\end{equation}
While Eq.~\eqref{eq:HCXLOC} is used to provide hard constraints, some additional points on each segment can be chosen, and continuity of the flux function in these points can be used as soft constraints. These constraints can be described as a set of linear equations as well; therefore, the unknown coefficients can be evaluated by standard least square methods.

\section{Plasma boundary reconstruction via Physic Informed Extreme Learning Machines}\label{section:PINNReconstruction}
This section introduces the main contribution of this paper, namely a boundary reconstruction algorithm based on a physics-informed ELM network. Before describing in detail our algorithm, a brief introduction to ELM networks is given. The interested reader can refer to~\cite{wang2022review} for more details.

\subsection{Extreme Learning Machines}
\begin{figure}[h!]
    \centering
    \includegraphics[width=0.4\linewidth]{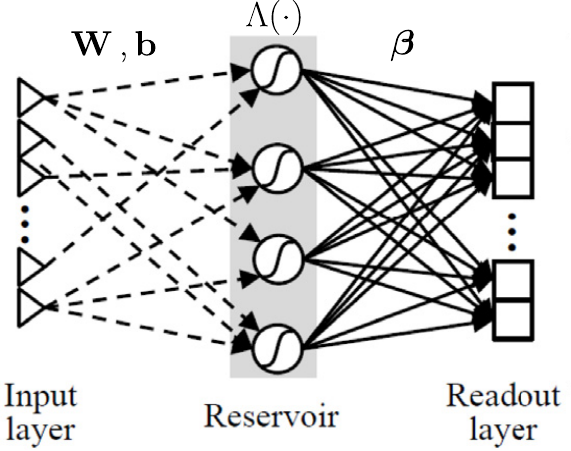}
    \caption{The basic ELM is made up of the reservoir and the output layer. The reservoir consists of nonlinear neurons randomly connected to the network input, while the output layer is connected to the nonlinear neurons through the weights to be optimized during the network's training.}
    \label{fig:ELMStruct}
\end{figure}

The basic architecture of an ELM network is reported in Fig.~\ref{fig:ELMStruct}. 

ELM training differs from traditional gradient-based learning methods 
because of the random initialization of the weights between the input and the hidden layer, as well as the biases of the hidden neurons. These parameters remain unchanged during training. The use of nonlinear activation functions in the hidden layer introduces the necessary nonlinearity into the network. With these fixed random values, the learning process reduces to solving a linear system. In fact, only the output weights (between the hidden and output layers) are learned. Hence, in the case of a single output network, as the one proposed in this paper to reconstruct the poloidal flux~$\psi(r,z)$ (see Section~\ref{sec3:ELM}), the~ELM architecture can be expressed by:
\begin{equation*}
    \mathcal{F}(\mathbf{x},\boldsymbol{\beta}) = \boldsymbol{\beta}^T \cdot\Lambda(\mathbf{W\cdot x}+\mathbf{b})\,,
\end{equation*}
where $\mathbf{W}\in\mathbb{R}^{\omega\times h}$ and $\mathbf{b}\in\mathbb{R}^\omega$ are randomly initialized, being $h$ and $\omega$ the number of network's inputs and neurons, respectively. The vector $\mathbf{x}\in\mathbb{R}^h$ is the input vector, while~$\boldsymbol{\beta}\in\mathbb{R}^\omega$ is the vector of unknown output weights. Note that in general, depending on the network output, $\boldsymbol{\beta}$ could be a two-dimensional matrix. The function~$\Lambda(\cdot)$ is the activation function, and can be chosen as any of the commonly used activation functions~(\texttt{tanh}, \texttt{sigmoid}, \texttt{logsig}, etc.), as long as they are twice differentiable in order to set the physics-informed constraints corresponding to~\eqref{eq:FluxFieldRel} and~\eqref{eq:GDeqHomo}.

As preliminary discussed in Section~\ref{section:Introduction}, the ELM network trains much faster than conventional~NNs, since it avoids iterative optimization. Despite the randomness, it retains the universal approximation capability, and theoretical analyses suggest that it has a higher potential to reach a globally optimal solution, if compared to traditional methods that train all parameters. These networks have been widely used for clustering, object recognition, and feature learning. Recently, some studies also showed how to use~ELMs to solve Partial Differential Equations with applications to Lyapunov's stability~\cite{zhou2024physics}. Therefore, here it is proposed to use an ELM to solve the homogeneous GS equation, which gives the poloidal flux function in the vacuum.

\subsection{Plasma boundary reconstruction via Physics-Informed ELMs}\label{sec3:ELM}

Let~$\psi_{ELM}(\mathbf{x},\boldsymbol{\beta})$ be the physics-informed ELM network used for the boundary reconstruction. For this application, the input vector consists of the coordinates of a generic point in the poloidal plane, i.e.,~$\mathbf{x} = \begin{pmatrix} r & z \end{pmatrix}^T$, while the output is a scalar value corresponding to the poloidal flux at~$\mathbf{x}$.  
Given the definition of the~$\Delta^*$ operator reported in Eq.~\eqref{eq:NablaStar}, by applying it to~$\psi_{ELM}(\mathbf{x},\boldsymbol{\beta})$, the homogeneous~GS equation is linear with respect to the parameters vector $\boldsymbol{\beta}$, yielding 
\begin{equation}\label{eq:GShomoELM}
    \boldsymbol{\beta}^T\cdot D_{GS}(\mathbf{x}) = 0\,,
\end{equation} where~$D_{GS}(\mathbf{x})\in\mathbb{R}^{\omega}$ represents $\Delta^* \Lambda(W\mathbf{x}+\mathbf{b})$. Therefore, once the required derivatives have been computed and enough points in the vacuum chamber are chosen, exploiting Eq.~\eqref{eq:GShomoELM}, a homogeneous linear system can be put in place to satisfy the homogeneous GS equation. Considering also the sensor readings, the physics-informed ELM can be trained to reconstruct the poloidal flux function by minimizing the following loss function:
\begin{equation}\label{eq:LossELM1}
    \text{Loss}(\boldsymbol{\beta}) = \frac{1}{N}\sum_{i=1}^N\lambda_0\cdot ||\boldsymbol{\beta}^T\cdot D_{GS}(\mathbf{x}_i)||^2+\lambda_1\cdot ||\boldsymbol{\beta}^T \cdot\Lambda(\mathbf{W\cdot x_\psi}+\mathbf{b}) - \mathbf{m_\psi}||^2 + \lambda_2\cdot ||\boldsymbol{\beta}^T\cdot D_\mathbf{B}(\mathbf{x}_\mathbf{B})-\mathbf{m_B}||^2\,,
\end{equation}
where: \begin{itemize}
    \item $D_\mathbf{B}(\mathbf{x_B})\in\mathbb{R}^{\omega\times n_\mathbf{B}}$ takes into account the derivatives~\eqref{eq:FluxFieldRel} evaluated at magnetic field sensors points~$\mathbf{x}_\mathbf{B}$;
    \item $\mathbf{m}_\psi$ is the vector of poloidal flux measures at points~$\mathbf{x}_\psi$;
    \item $\mathbf{m}_\mathbf{B}$ is the vector of poloidal magnetic field measures at points $\mathbf{x}_\mathbf{B}$;
    \item $\mathbf{x}_i\in\mathbb{R}^{g}$ are the collocation points placed in the poloidal plane to satisfy Eq.~\eqref{eq:GShomoELM};
    \item $\lambda_i$, with~$i=0,1,2$, are weights for the different components of the loss function~\eqref{eq:LossELM1}.
\end{itemize} 
Since~\eqref{eq:LossELM1} is a quadratic function of $\boldsymbol{\beta}$, the unknown set of weights can be determined by standard least squares methods. Hence, once the optimal vector $\boldsymbol{\beta}^*$ is computed, $\psi_{ELM}(\mathbf{x},\boldsymbol{\beta}^*)$ represents an approximation of the poloidal flux function and can be used to identify the plasma boundary. Due to its simplicity, the neural network can be trained every time new sensor data is available, making it suitable for real-time implementation. It follows that, given the loss function~\eqref{eq:LossELM1}, the proposed physics-informed~ELM network approximates the solution of the~GS PDE. 

The solution of the homogeneous~GS equation can also be imposed as a hard constraint by choosing $\boldsymbol{\beta}$ in the null-space of~$D_{GS}(\mathbf{X}_c)^T\in \mathbb{R}^{N\times\omega}$, being~$\mathbf{X}_c~\in~\mathbb{R}^{N\times2}$ the matrix formed by the $N$ (with $N<\omega$) collocation points $\mathbf{x}_i$. Let~$E~\in~\mathbb{R}^{\omega\times\left(\omega-\rank\left(D_{GS}(\mathbf{X}_c)\right)\right)}$ be a matrix such that \begin{equation*}
    \begin{cases}
    D_{GS}(\mathbf{X}_c)^TE=0\\
    E^TE=I
\end{cases}\,.
\end{equation*}
Therefore, it represents an orthonormal basis in the null-space of $D_{GS}(\mathbf{X}_c)^T$. By choosing $  \boldsymbol{\beta}=E\Tilde{\boldsymbol{\beta}}$, then Eq.~\eqref{eq:GShomoELM} is automatically satisfied at points $\mathbf{x}_i$. As a consequence, the choice of the collocation points becomes crucial. In fact, to have a good reconstruction of the plasma shape, they need to be placed in the vacuum vessel where Eq.~\eqref{eq:GDeqHomo} is always satisfied. One possibility is to place the points $\mathbf{x}_i$ along the first wall (or in its near proximity).

Note that $\Tilde{\boldsymbol{\beta}}\in\mathbb{R}^{\left(\omega-\rank\left(D_{GS}(\mathbf{x}_c)\right)\right)}$, reducing the number of free parameters to be optimized. On the other hand, the loss function to be minimized with the training needs now to take into account only the flux and magnetic field  measurements:
\begin{equation}
\label{eq:LossELM2}
    \text{Loss}(\boldsymbol{\beta}) = \frac{1}{N}\sum_{i=1}^N\lambda_1\cdot|\boldsymbol{\Tilde{\beta}}^T \cdot E^T\cdot\Lambda(\mathbf{W\cdot x_\psi}+\mathbf{b}) - \mathbf{m_\psi}||^2  +\lambda_2\cdot||\Tilde{\boldsymbol{\beta}}^T\cdot E^T\cdot D_B(\mathbf{x}_B)-\mathbf{m_B}||^2\,.
\end{equation}

Finally, in order to properly identify the neural network weights, the number of degrees of freedom $\omega-\rank\left(D_{GS}\left(\mathbf{x}_c\right)\right)$ needs to be less than or equal to the number of available measurements~$n_m$.

As anticipated in the Introduction, the main advantage of this method over the others presented in the literature lies in its simplicity. The nonlinearities introduced by the~NN activation functions allow to estimate accurately the whole poloidal flux function, without using local approximations, as the~XLOC algorithm. Specifically, the minimization of the loss function~\eqref{eq:LossELM2} requires only a single matrix pseudo-inversion that is performed offline, while in real-time, a single matrix multiplication is sufficient at each sampling period to reconstruct the flux map in the vacuum region. An estimation of the poloidal magnetic field in the vacuum can be given by exploiting Eq.~\eqref{eq:FluxFieldRel}. 

In contrast, other approaches typically involve either locally approximating the poloidal flux function and then putting together the results, or identifying current sources and summing their contributions. With the latter methods, accounting for the eddy currents is a challenging operation. Indeed, current sources need to be placed inside the whole vessel. On the other hand, the proposed PINN-based method, similarly to the~XLOC-like algorithm, directly approximates the poloidal flux function using a set of basis functions. Therefore, the passive currents' contribution is directly taken into account.
\begin{figure}[!thb]
    \centering
    \includegraphics[width=0.5\linewidth]{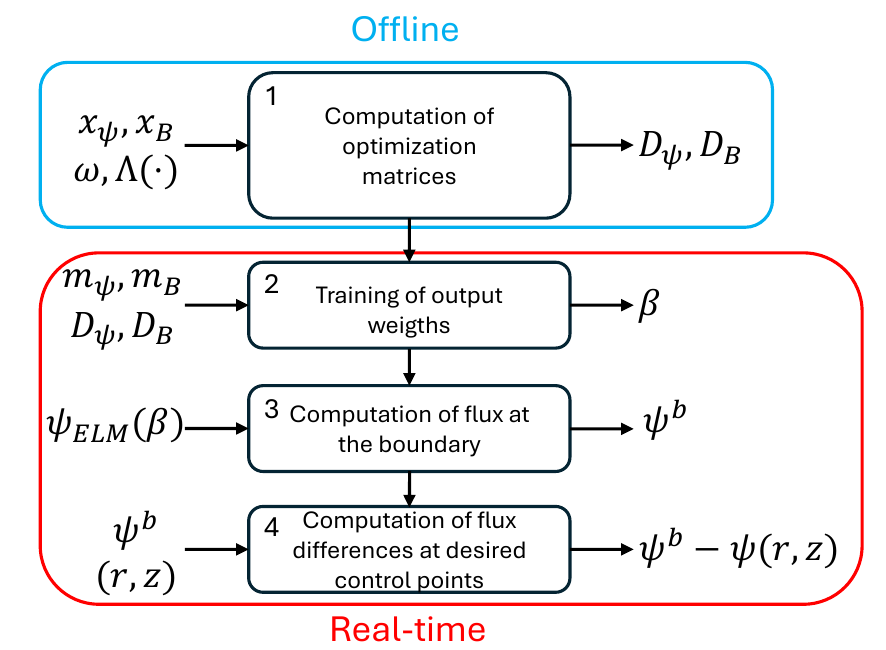}
    \caption{Simplified scheme of the CARONTE algorithm workflow.}
    \label{fig:CARONTE}
\end{figure}

 Moreover, many of these methods depend on a fixed grid of points. Instead, the proposed approach allows for the evaluation of the poloidal flux function at arbitrary points once the parameters have been identified, giving a continuous approximation of the poloidal flux function. 
Finally, Fig.~\ref{fig:CARONTE} shows the workflow of CARONTE. Once the hyper-parameters of the physics-informed ELM network are set, i.e.~the number of neurons and the corresponding activation functions, as well as the randomly chosen input weights and biases, the~CARONTE algorithm consists of the following steps:
\begin{enumerate}
\item[i)] offline computation of the matrices that allows to compute the optimal output weights given the actual value of the magnetic measurements;
\item[ii)] the online training of the network output weights at each sampling time;
\item[iii)] the computation of the flux at the plasma boundary;
\item[iv)] the computation of the flux at the desired control points, by evaluating the trained ELM NN at the corresponding coordinates.
\end{enumerate}
Note that steps ii) and iv) are simple matrix multiplications, and the evaluation of the estimated flux at a given point can be done straightforwardly in one single operation. On the other hand, Step iii) is required by any reconstruction algorithm used for control purposes.

\begin{figure*}[h!]
    \centering
    \includegraphics[width=0.49\linewidth]{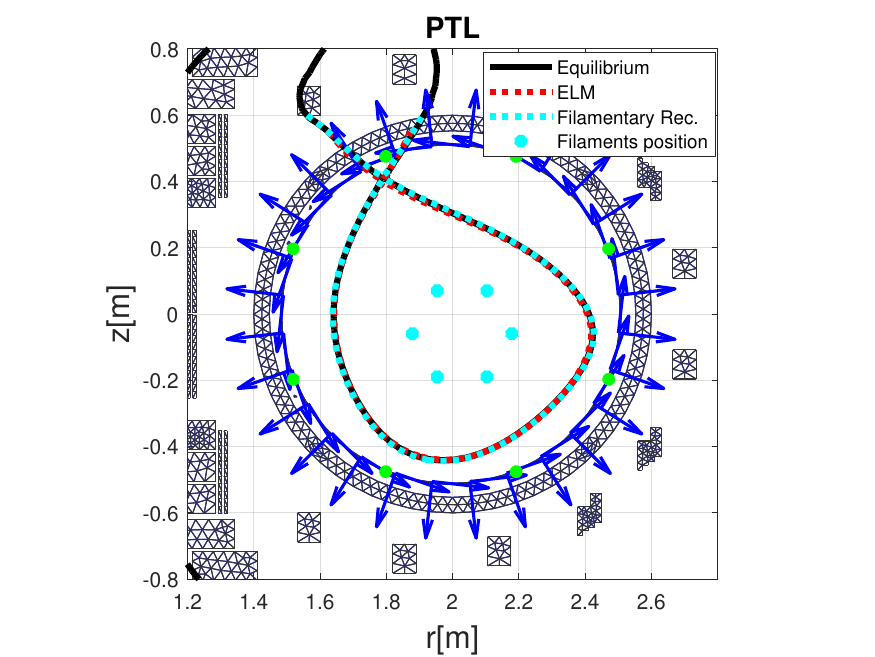}
    \includegraphics[width=0.49\linewidth]{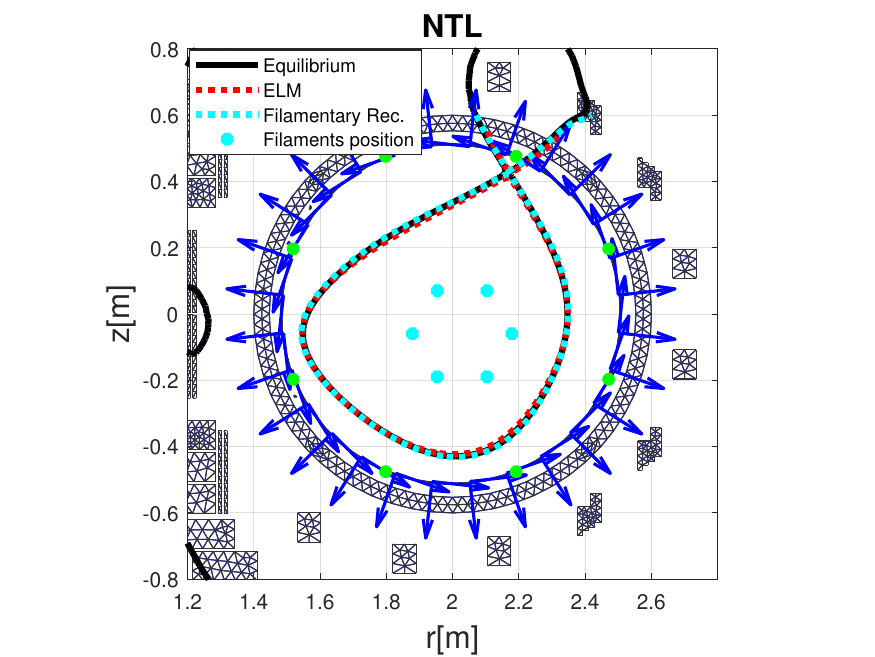}
    \caption{These figures show the plasma equilibrium computed by the CREATE code and the reconstruction with the filamentary method~(cyan) and with the physics-informed~ELM network (red). On the left, the~PTL configuration and on the right the~NTL configuration. The agreement with the equilibrium is very good with both reconstruction methods. The green dots represent the flux loops, the arrows represent the magnetic probes.}
    \label{fig:PTLNTL}
\end{figure*}

\section{Validation}\label{section:validation}

In this section, the proposed method is validated using~RFX-mod2 and~DTT plasma equilibria. RFX-mod2 is a small circular device, while DTT is under construction and, once completed, will be one of the most relevant superconducting tokamak machines.
The proposed method is proven to be effective in both cases, showing that the PINN manages to generalize well the plasma shape disregarding of a particular machine or equilibrium. Furthermore, the studies presented in this section will highlight the effects of different activation functions and of the initialization of the random input weights and biases. The considered plasma equilibria are obtained with the CREATE-L~\cite{albanese1998linearized} and CREATE-NL~\cite{albanese2015create} codes. These codes have been extensively validated and they effectively approximate the plasma equilibria, giving reliable values of the magnetic sensors' readings. All the test cases shown in this section consider static plasma equilibria, i.e., no passive currents are present in the conducting structures surrounding the plasma. The quality of the reconstructed shape is evaluated by comparing it with the equilibrium one given by the~CREATE codes, which is considered as ground truth. For the~DTT case, the values of some gaps critical for control are computed and then compared to the equilibrium ones. Moreover, to assess the quality of the overall reconstruction, using a larger number of gaps with respect to the typical number of controlled ones, the root mean square deviation~(RMSD) with respect to the equilibrium is computed. 

\subsection{Test cases}\label{subsection:testcases}
The considered test cases aim at testing the generalization capabilities of the~PINN and to test if it can be applied to any machine and plasma equilibrium. While doing so, we test two different activation functions, showing how much this choice matters. In particular, the logistic sigmoid (\texttt{logsig}) and the hyperbolic tangent (\texttt{tanh}) are used. Moreover, it is shown that a random choice of the network input weights and biases permits to reconstruct the plasma boundary. However, the performance and generalization capabilities of the network can be greatly improved by properly choosing the random distribution for these network parameters, as it will be shown in the second test case. However, the possibility of further improving the performance by a specific choice of the weights and biases, and of the activation functions, is left for future studies. In what follows, the~CARONTE reconstructions of the~RFX-mod2 equilibria are compared with those obtained using the filamentary reconstruction algorithm introduced in Section~\ref{subsection:filamentary}, which is known to be very effective for small circular machines. Regarding the~DTT equilibria, the CARONTE reconstructions are compared with those of the~XLOC-like code. Robustness against noise is also evaluated in this case.  
\begin{figure}[h]
    \centering
    \includegraphics[width=0.4\linewidth]{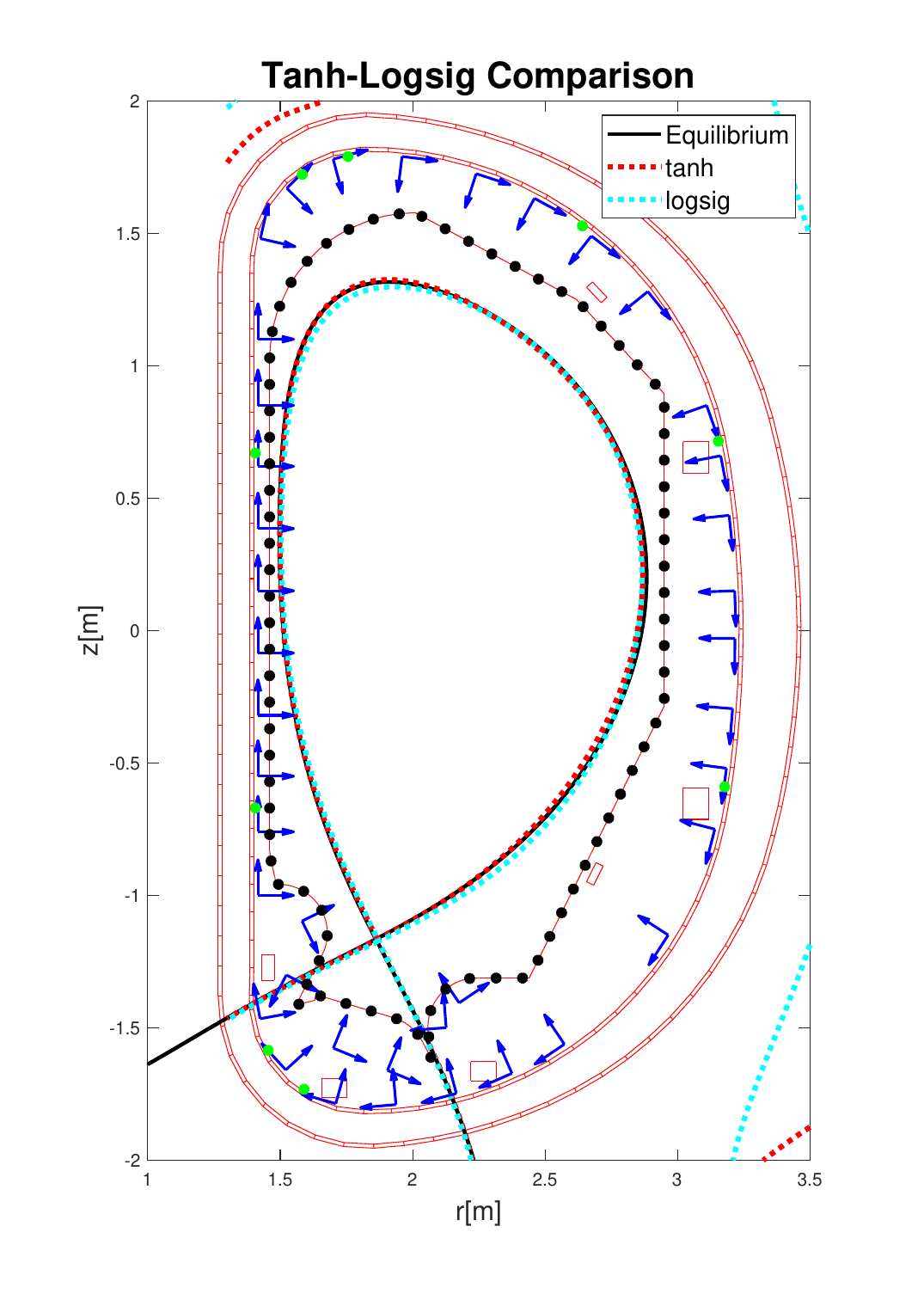}
    \caption{Comparison between reconstruction using \texttt{tanh} (red) or \texttt{logsig} (cyan) functions. In blue, the magnetic field sensors, and the green dots are the flux loops. The black dots represent the points at which the homogeneous GS condition is imposed.}
    \label{fig:TanhLogSig}
\end{figure}
\begin{figure*}[h]
    \centering
    \includegraphics[width=0.49\linewidth]{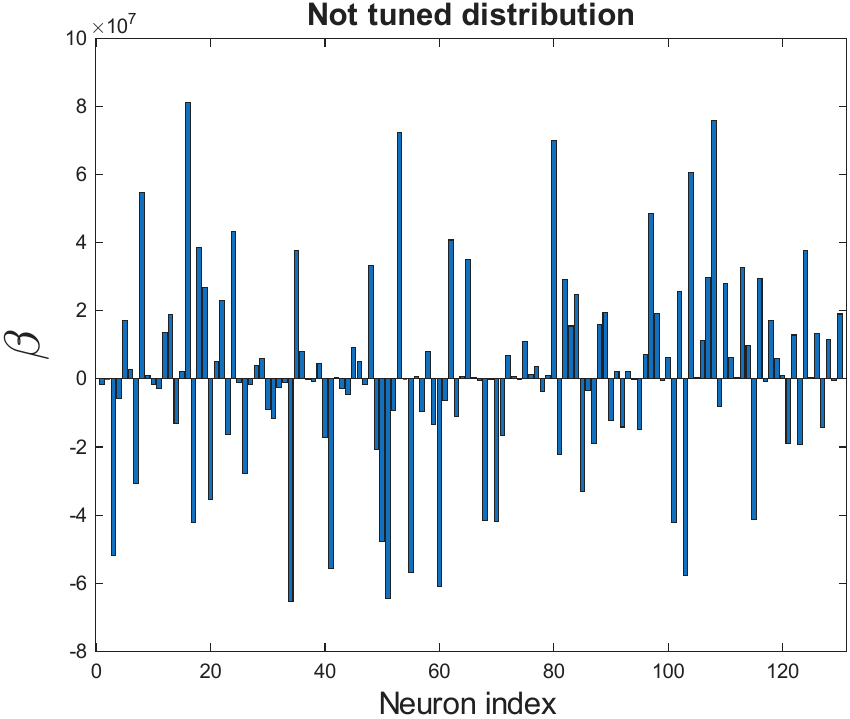}
    \includegraphics[width=0.49\linewidth]{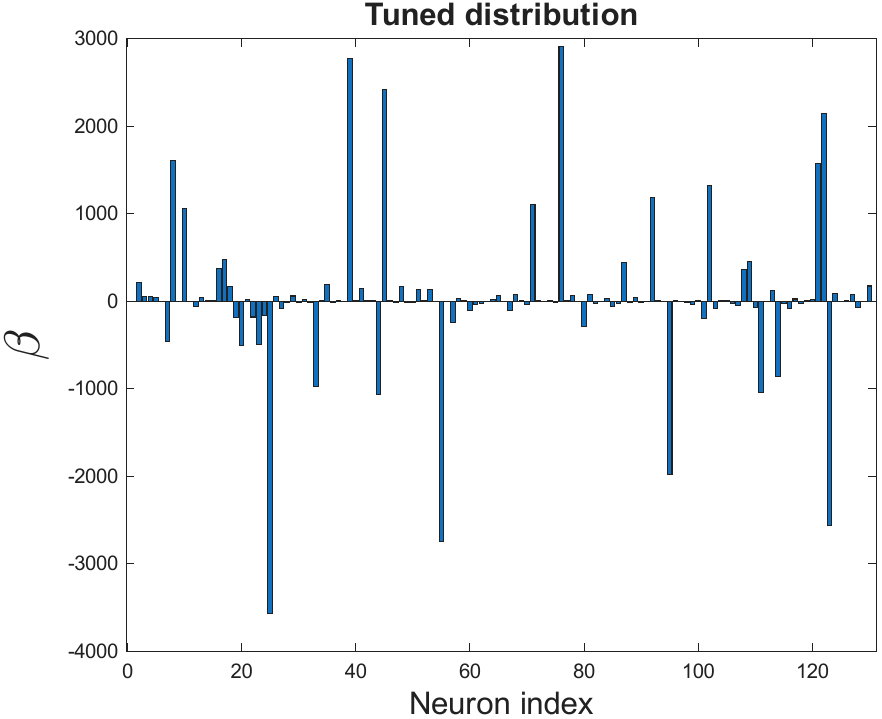}
    \caption{On the left, the resulting output weights~$\boldsymbol{\beta}$ of the physics-informed ELM network, when the input weights and biases are randomly generated Gaussian distributions centered at the origin and with amplitude equal to~1. In this case, the order of magnitude for the elements of~$\beta$ is~$10^7$, which is due to matrices that are ill-conditioned. By choosing the random distribution for the input weights and biases as described in Section~\ref{subsection:testcase2}, the maximum value for the~$\beta$ components is in the order of $10^3$, due to a better conditioning of the optimization matrices.}
    \label{fig:OptAlpha}
\end{figure*}
\begin{figure*}[h]
    \centering
    \includegraphics[width=0.32\linewidth]{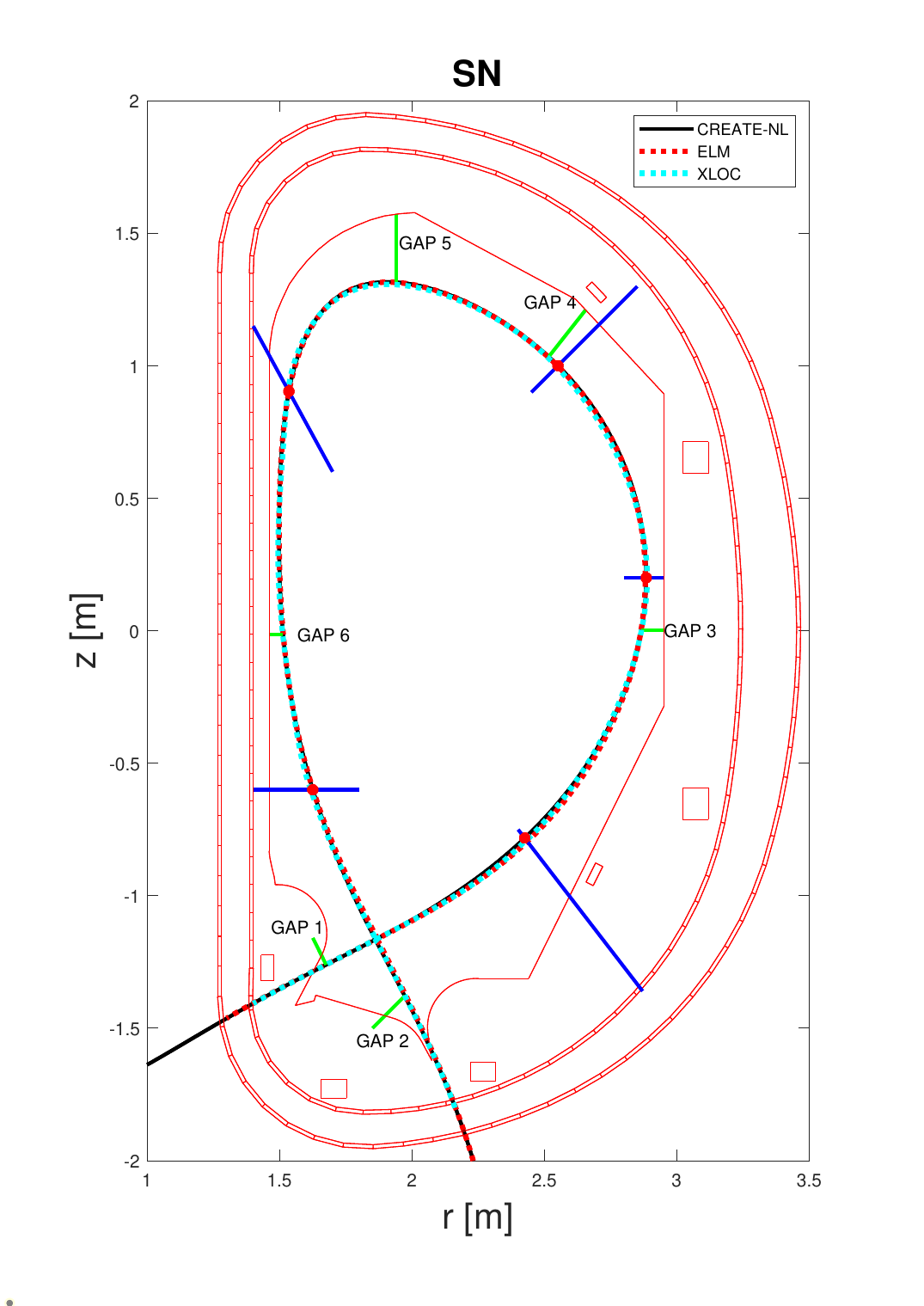}
    \includegraphics[width=0.32\linewidth]{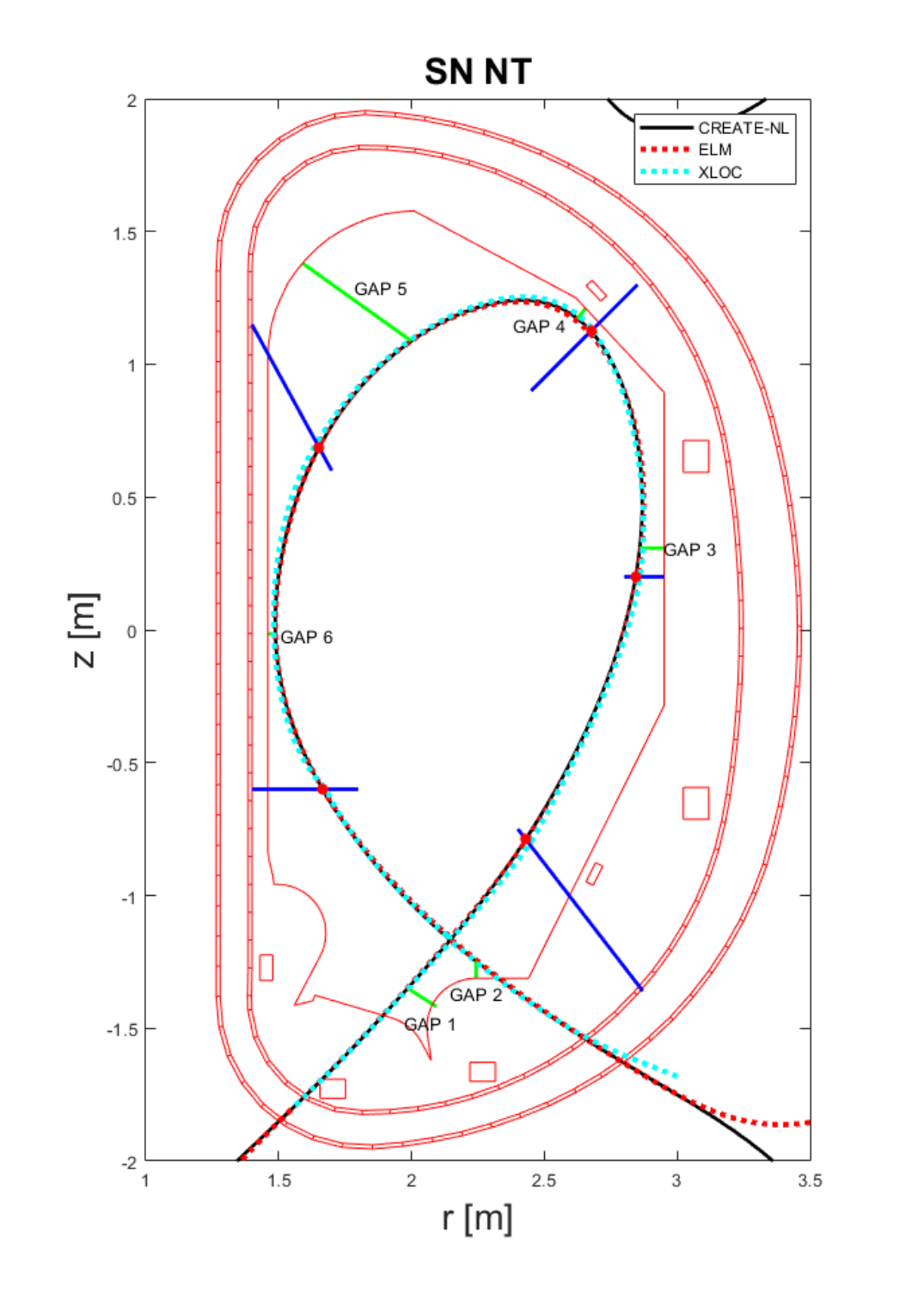}
    \includegraphics[width=0.32 \linewidth]{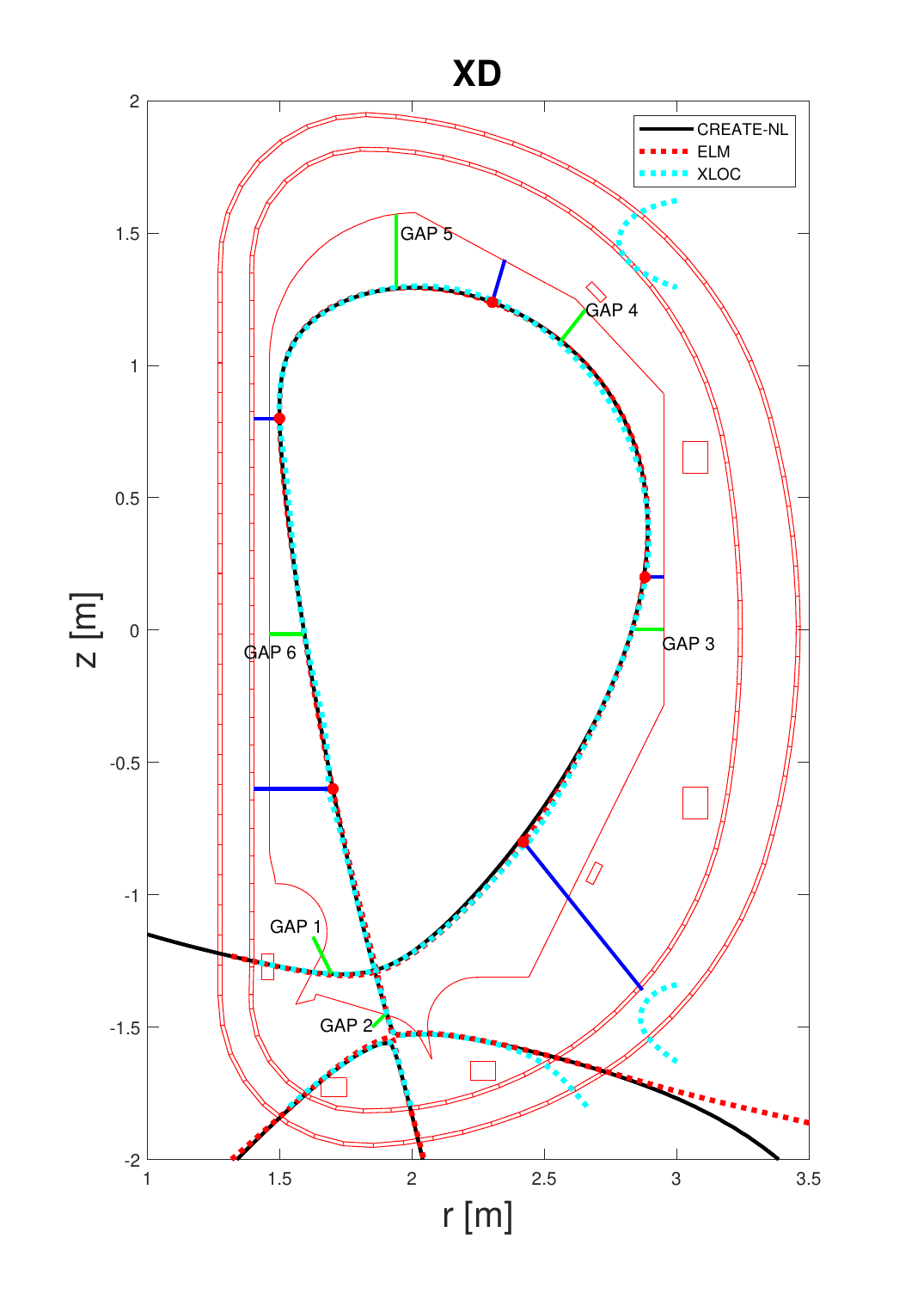}

    \caption{These figures show the plasma equilibrium computed by the~CREATE equilibrium code, the reconstruction with the~XLOC-like approach~(cyan) and with the physics-informed~ELM network (red). On the left, the SN configuration, in the center the SN-NT configuration, and on the right the~XD configuration. The agreement with the equilibrium is good with both reconstruction methods, but the ELM shows superior performance. The green dots represent the flux loops, the arrows represent the magnetic probes. The points at which the homogeneous~GS is imposed are shown in Fig.~\ref{fig:TanhLogSig}.}
    \label{fig:ELMXLOCComp}
\end{figure*}
\subsection{Test case \#1 - RFX-mod2}\label{subsection:testcase1}

In this section, two different RFX-mod2 plasma configurations are considered. The available magnetic sensors in~RFX-mod2 are 8 flux-loops and 56 magnetic probes, including both normal and tangential ones,~(see~Fig.~\ref{fig:PTLNTL}). For these tests, the logistic sigmoid is used as an activation function:
\[ \Lambda(x) = \frac{1}{1+e^{-x}}\,.\]
The number of neurons used in this case is~$\omega=80$. To impose~Eq.~\eqref{eq:GShomoELM} as a hard constraint, 24 points along the first-wall are chosen, leaving~$\omega-\rank\left(D_{GS}\left(\mathbf{x}_c\right)\right)=56$ degrees of freedom to the~ELM to match the magnetic measurements. The random weights~$\mathbf{W}$ and $\mathbf{b}$ are chosen with a uniform Gaussian distribution whose values are contained in the interval \([-0.5\quad0.5]\).  Finally, since the order of magnitudes between magnetic flux and field are different, weighting terms, i.e., the~$\lambda_i$ in~\eqref{eq:LossELM2}, are chosen to better fit the magnetics. 

For the filamentary reconstruction, 6 filaments are used to describe the current distribution inside the plasma. They are placed on a circumference, with radius equal to~15~cm, centered on the plasma current centroid position. 
Note that, other than the magnetic measurements, the filamentary method also needs the measurements of the active currents flowing in the PF coils, hence it is sensitive also to the corresponding measurement noise, while the CARONTE approach does not need this additional information.

Fig.~\ref{fig:PTLNTL} shows the reconstruction with both methods for the Positive Triangularity L-mode~(PTL) configuration and the Negative Triangularity L-mode~(NTL). Although the two configurations are completely different, without any change in the network's tuning, both algorithms manage to effectively reconstruct the plasma shape. Tab.~\ref{tab:RMSDRFX} shows the performance in terms of~RMSD between the reconstructed shapes and the equilibrium one. The RMSD has been computed by computing the error along~25 gaps whose segments connect several points on the first wall with the center of the chamber. These results are in line with what is reported in~\cite{kudlacek2015real}, where the authors obtained reconstruction errors of about~10~mm with respect to the~RFX-mod equilibrium computed with the~MAXFEA code~\cite{barabaschi1993maxfea}. 

Being RFX-mod2 a very small machine, with circular symmetry and small plasmas, the filaments placed in the plasma center are very effective in reconstructing the plasma shape. However, this simple approach finds its limits in larger machines with elongated plasmas. In fact, the distribution of the filaments should then be adapted to the evolution of the size and shape of the plasma. Nevertheless, we proved that with a completely random choice of the parameters $\mathbf{W}$ and $\mathbf{b}$ and with a~\texttt{logsig} function, the physics-informed ELM can accurately reconstruct the~LCFS on a small circular machine\linebreak like~RFX-mod2. Note also that although $ \Lambda(x)\in(0\quad1)$, given enough neurons, the ELM manages to accurately approximate the poloidal flux function.

\begin{table}[]
    \centering
    \begin{tabular}{c|c|c}
    \hline
         &    \multicolumn{2}{c}{\large RFX-mod2 equilibrium}  \\
         \hline
        Reconstruction approach & PTL & NTL\\ 
        \hline
         CARONTE & 0.26 cm & 0.48 cm \\
        Filamentary model & 0.27 cm& 0.24 cm\\
        \hline
    \end{tabular}
    \caption{RMSD in cm of reconstructed shapes with respect to~CREATE equilibrium. The RMSD is below~1~cm in each case, therefore these errors are negligible.}
    \label{tab:RMSDRFX}
\end{table}
\subsection{Test case \#2 - DTT}\label{subsection:testcase2}
This test case deals with the advanced tokamak scenario of the~DTT device. Similarly to RFX-mod2, DTT~is highly flexible and capable of sustaining single null plasma equilibria with both positive~(SN) and negative\linebreak triangularity~(SN-NT), as well as X-Divertor~(XD) configurations featuring two X-points. This flexibility motivates the need for a fast and robust algorithm capable of accurately identifying the~LCFS. Moreover, given the high energies involved, precise reconstruction of the plasma boundary is even more critical than in the previous case. Indeed, as reported in~\cite{fiorenza2026preliminary}, the accepted accuracy for these large-scale machines is~$1~\mathrm{cm}$, which represents a strict requirement. At the end of this section, we will demonstrate that our method satisfies this criterion.
\begin{table*}[h!]
\centering
\begin{tabular}{c|c|c|c|c|c|c|c|c|c|c|c|c}
\hline
  \multicolumn{1}{c}{} & \multicolumn{6}{c|}{XLOC-like} & \multicolumn{6}{c}{CARONTE}  \\
\hline
  \multicolumn{1}{c}{} & \multicolumn{2}{|c|}{SN} & \multicolumn{2}{c|}{SN-NT} & \multicolumn{2}{c|}{XD} & \multicolumn{2}{c|}{SN} & \multicolumn{2}{c|}{SN-NT} & \multicolumn{2}{c}{XD}  \\
\hline
 Gaps&  $\mu_e $ & $\sigma_e $  &  $\mu_e $ & $\sigma_e $ &  $\mu_e $ & $\sigma_e $  &  $\mu_e $ & $\sigma_e $&  $\mu_e $ & $\sigma_e $  &  $\mu_e $ & $\sigma_e $\\
\hline
1 &  0.08 & 0.10  & -0.37 & 0.20 &  0.96 & 0.30  &  0.05 & 0.12 & -0.15 & 0.19 & -0.47 & 0.33 \\
2 & -0.01 & 0.09  & -0.03 & 0.40 & -0.70 & 0.83  & -0.46 & 0.11 & -0.53 & 0.18 & -0.12 & 1.15 \\
3 & -0.37 & 0.63  &  0.13 & 0.50 & -0.20 & 0.73  & -0.18 & 0.24 &  0.59 & 0.17 &  0.15 & 0.27 \\
4 & -0.50 & 0.74  &  0.62 & 1.27 &  0.17 & 1.07  & -0.18 & 0.15 & -0.93 & 0.28 & -0.25 & 0.12 \\
5 & -0.02 & 1.33  &  0.72 & 1.03 &  1.29 & 1.03  & -0.08 & 0.28 &  0.07 & 0.24 &  0.14 & 0.18 \\
6 &  0.75 & 0.56  &  0.75 & 0.81 &  0.14 & 0.48  & -0.05 & 0.08 &  0.02 & 0.13 &  0.02 & 0.07 \\
\hline
RMSD &  0.96 & 0.28  &  1.42 & 0.52 &  1.63 & 0.44  &  0.71 & 0.15  & 0.55 & 0.09 & 0.82 & 0.19\\
\hline
\end{tabular}
\caption{Comparison between plasma boundary reconstruction of DTT plasmas carried out with the XLOC-like algorithm and the physics-informed ELM network. For both approaches, this table reports the error on the gap values with respect to the corresponding equilibrium computed with the CREATE codes. The gaps used for this evaluation are shown in green in Figure~\ref{fig:ELMXLOCComp}. All the values reported here are expressed in cm.}
\label{tab:ELMXLOCComp}
\end{table*}
First, using a~SN equilibrium, we show that the \texttt{logsig} activation function, combined with the same random distribution used in the previous test case for the input weights and biases, still yields an acceptable reconstruction of the flux map. Since the~DTT device is significantly larger, and a higher number of magnetic measurements is available, the number of neurons used in this case is set equal to~$\omega=130$. Indeed, referring to the magnetic sensor layout of~DTT reported in Fig.~\ref{fig:TanhLogSig}, 40~biaxial magnetic field sensors returning~80 measurements of both the tangential and normal field components, and 9 flux loops are available. Along the first wall, 59 points are selected to impose the~GS equation in the vacuum region, resulting in a total of~71 degrees of freedom for the physics-informed~ELM network.

As in the previous case, also for this more complex configuration, the network reconstructs the~LCFS. However, the resulting~RMSD, computed as described in the previous section, is~$1.6~\mathrm{cm}$, which is close to the~$1~\mathrm{cm}$ requirement. Therefore, a further improvement is needed, and can be obtained by simply changing the activation function, while keeping all other network's hyper-parameters unchanged, even the random ones. Hence by choosing
\[
 \Lambda(x) = \texttt{tanh}(x),
\]
the RMSD decreases to $0.6~\mathrm{cm}$, fully satisfying the accuracy requirement. The two reconstructed shapes are shown in Fig.~\ref{fig:TanhLogSig}. Furthermore, since~$\Lambda(x) \in (-1,\,1)$, each neuron carries higher information content, allowing for a reduction in the number of degrees of freedom required by the~ELM. Indeed, comparable performance is also achieved with only~48 degrees of freedom, hence increasing the points for the homogeneous GS hard-constraints up to~82. The location of these~82 points is reported in~Fig.~\ref{fig:TanhLogSig}. 

The conditioning of the $D_B(\mathbf{x_B})$ and~$\Lambda(\mathbf{Wx_\psi}+\mathbf{b})$ matrices strongly depend on the parameters~$\mathbf{W}$ and~$\mathbf{b}$. If~these two matrices are not well-conditioned, the resulting~$\boldsymbol{\beta}$ weights can assume values in a very large range, which may also affect the robustness against noise. If the centers of the activation functions are better scattered on the poloidal plane, the entries of these matrices will result in more heterogeneous values and hence are better conditioned. 
The weights and biases generated by such tuned random distributions permit to obtain much better conditioned matrices, therefore smaller output weights~$\boldsymbol{\beta}$. 
Note that this tuning process depends on the position of the sensors. Therefore, the optimization of the random weights can be done based on the machine geometry (and on the size of the network), and it is independent of the particular plasma configuration to be achieved. In Fig.~\ref{fig:OptAlpha} it is shown how the choice of $\mathbf{W}$ and $\mathbf{b}$ can affect $\boldsymbol{\beta}$. In particular, in the tuned case, the activation functions have been translated in a point of the poloidal plane $\mathbf{x}_0 = [2.1\quad0]^T$, while $\mathbf{W}$ and $\boldsymbol{\beta}$ are chosen with a gaussian distribution contained in the interval \([-1.5\quad1.5]\). 

We now compare the performance of the proposed method with that of XLOC-like, using three entirely different plasma configurations. In addition to evaluating the~RMSD with respect to the equilibria computed with the CREATE codes, we assess the quality of reconstruction along the critical gaps highlighted in green in Fig.~\ref{fig:ELMXLOCComp}, which were the same as those chosen in~\cite{opereq}, where the performance of the XLOC-like algorithm on~DTT has been evaluated.

With 48 degrees of freedom and tuned distribution for the random weights, the proposed approach can accurately estimate the three different plasma boundaries without modifying the network hyper-parameter. In Fig.~\ref{fig:ELMXLOCComp}, the shapes reconstructed by the two algorithms are shown. Although the performance of the XLOC-like algorithm is also somewhat satisfactory, it is challenging to reconstruct the plasma shape of these three equilibria using the same algorithm setup. This implies that for different plasma configurations, retuning of the algorithm may be necessary, including adjustments to the number of regions, their positions, and the number of sensors used per region. Indeed, in Fig.~\ref{fig:ELMXLOCComp}, the blue lines represent the segments defining the different regions considered by the XLOC-like algorithm, while the red dots indicate the points where continuity of the different polynomials is imposed, according to Eq.~\ref{eq:HCXLOC}. When switching from the two SN equilibria to the~XD configuration, both the regions and sensor weights require modification, whereas such changes are unnecessary for~CARONTE reconstruction.

To systematically assess the performance of these two algorithms, noise is added to the magnetic field measurements. The total measurement error consists of both a relative and an absolute term. Specifically, it is assumed that
\[
m = m_0 (1 + n_r) + n_a\,,
\]
where
\begin{itemize}
  \item \(m_0\) is the ideal (noise-free) value measured by a sensor;
  \item \(m\) is the actual measured value;
  \item \(n_r\) is the relative error, modeled as a zero-mean Gaussian random variable with standard deviation \(\sigma_r\);
  \item \(n_a\) is the absolute error, modeled as a zero-mean Gaussian random variable with standard deviation~\(\sigma_a\).
\end{itemize}
The parameters \(\sigma_a\) and \(\sigma_r\) are selected to realistically model sensor positioning errors and misalignments. For~DTT, these are estimated as \(\sigma_r = 0.00079\)\linebreak and \(\sigma_a = 0.64\,\text{mT}\). In the analyses presented here, more conservative values of \(\sigma_r = 0.003\) and \(\sigma_a = 1.5\,\text{mT}\) were used.

After 500 noise realizations, the reconstruction errors of the chosen gaps are summarized in Tab.~\ref{tab:ELMXLOCComp}, reporting the mean \((\mu_e)\) and standard deviation \((\sigma_e)\) in centimeters. These results clearly demonstrate the superior performance of the ELM reconstruction. Along the defined gaps, although the requirement is met on the mean, considering the resulting standard deviation, the XLOC-like algorithm frequently fails to meet the 1~cm accuracy requirement. On the other hand, the CARONTE algorithm satisfies the accuracy criterion at a 95\% confidence level in all but very few cases. Nevertheless, in these exceptional cases, the mean error remains within the acceptable limit anyway.

\section{Conclusive remarks and future perspective}\label{section:Conclusions}
In this work, we propose a PINN-based plasma boundary reconstruction algorithm suitable for real-time deployment. The~NN adopted to solve the homogeneous~GS equation is an ELM. Thanks to its simple structure, this~NN allows real-time training of its parameters, dynamically adjusting its output weights according to the available magnetic sensor readings and, consequently, to the plasma equilibria reached during the pulse. 

First, we demonstrated that even with suboptimal network parameters and activation functions, for small-sized and circular machines such as RFX-mod2, the algorithm effectively reconstructs the plasma boundary.
We then assessed the performance of the proposed approach on more complex configurations, considering~DTT scenarios and showing that with the same choices for the network parameters, the model still succeeds in reconstructing the boundary, although it does not meet the requested accuracy. However, with a proper and relatively straightforward network tuning, we show that the proposed algorithm can outperform established methods such as~XLOC. Specifically, the physics-informed ELM network generalizes the poloidal flux function more effectively since, unlike~XLOC, it does not require a retuning of the parameters when changing the reference equilibrium. Indeed, while the considered~XLOC-like algorithm achieves the 1\,cm accuracy requirement on average, our approach proves to be significantly more robust to noise, satisfying the accuracy criteria in almost all cases at a~95\% confidence level. This approach leverages the generalization capability of neural networks without requiring extensive, time-consuming training on experimental data, thereby enhancing the practicality of implementing NNs on these devices.
This is possible thanks to the simple structure of the~ELM networks, which allows a better understanding of the optimization process, and hence of its manipulation.

It is also worth remarking here that, if an isoflux plasma boundary control strategy is adopted~\cite{yuan2013plasma,de2024control}, the proposed network returns directly the flux at the control points, without the need of any interpolation on a grid. If the plasma boundary is controlled by regulating the gaps, then these must be estimated with a dichotomic search along segments, or by training a dedicated~NN; the latter is one of the possible future enhancements of the proposed approach.

Regarding future work, several research directions remain open. We have shown that a simple randomization of the initial weights can enhance algorithm performance. It should be possible to automate a procedure that optimizes the conditioning of the matrix based solely on sensor positions, regardless of the specific plasma equilibrium. Furthermore, the algorithm's performance should be tested in more realistic scenarios where strong passive currents are present. Finally, a complete real-time equilibrium reconstruction algorithm capable of solving Eq.~\ref{eq:GDsys} could be developed. These aspects are left for future investigation.


\section*{Acknowledgments}
The authors would like to thank Prof. Massimiliano Mattei.

\medskip

This work has been carried out within the framework of the EUROfusion Consortium, partially funded by the European Union via the Euratom Research and Training Programme (Grant Agreement No 101052200 — EUROfusion).

\medskip

The research leading to these results has been partially supported also by the Project ``TRAINER - Tokamak plasmas daTa-dRiven identificAtIon and magNEtic contRol" CUP~E53D23014670001 funded by EU in NextGenerationEU plan through the Italian ``Bando Prin 2022 - D.D.~1409 del~14-09-2022" by~Italian University and Research Ministry~MUR.

\medskip

The work of Federico Fiorenza has been carried out within the framework of the Italian National Recovery and Resilience Plan (NRRP), funded by the European Union - NextGenerationEU (Mission 4, Component 2, Investment~3.1~-~Area ESFRI Energy - Call for tender No. 3264 of 28- 12-2021 of~MUR, Project ID IR0000007, MUR Concession Decree No.~243 del 04/08/2022, CUP~B53C22003070006, ``NEFERTARI – New Equipment for Fusion Experimental Research and Technological Advancements with RFX Infrastructure"). 

\bibliographystyle{IEEEtran}
\bibliography{references}

@misc{ITER,
title = "{ITER website}",
howpublished = "https://www.iter.org/",
note = "accessed 12.11.2025",
}

@article{romanelli2024divertor,
  title={{Divertor Tokamak Test facility project: status of design and implementation}},
  author={Romanelli, Francesco and others},
  journal={Nuclear Fusion},
  volume={64},
  number={11},
  pages={112015},
  year={2024},
}

@article{kurihara2000new,
  title={{A new shape reproduction method based on the Cauchy-condition surface for real-time tokamak reactor control}},
  author={Kurihara, K},
  journal={Fusion Engineering Design},
  volume={51},
  pages={1049--1057},
  year={2000},
}

@techreport{opereq,
    author = "Vv.Aa.",
    title = "{{2023 modelling analysis on measurement capability progress of design activities}}",
    institution = "ENEA",
    year = "2023",
    note = "Available upon request to the corresponding author",
}

@article{bonotto2024reconstruction,
  title={Reconstruction of plasma equilibrium and separatrix using convolutional physics-informed neural operator},
  author={Bonotto, Matteo and Abate, Domenico and Pigatto, Leonardo},
  journal={Fusion Engineering and Design},
  volume={200},
  pages={114193},
  year={2024},
  publisher={Elsevier}
}

@article{wan2023machine,
  title={A machine-learning-based tool for last closed-flux surface reconstruction on tokamaks},
  author={Wan, Chenguang and Yu, Zhi and Pau, Alessandro and Sauter, Olivier and Liu, Xiaojuan and Yuan, Qiping and Li, Jiangang},
  journal={Nuclear Fusion},
  volume={63},
  number={5},
  pages={056019},
  year={2023},
  publisher={IOP Publishing}
}

@article{zhou2024physics,
  title={{Physics-informed extreme learning machine Lyapunov functions}},
  author={Zhou, Ruikun and Fitzsimmons, Maxwell and Meng, Yiming and Liu, Jun},
  journal={IEEE Control Systems Letters},
  volume={8},
  pages={1763--1768},
  year={2024},
  publisher={IEEE}
}

@article{wang2024neural,
  title={Neural-network-based free-boundary equilibrium solver to enable fast scenario simulations},
  author={Wang, Zibo and Song, Xiao and Rafiq, Tariq and Schuster, Eugenio},
  journal={IEEE Transactions on Plasma Science},
  volume={52},
  number={9},
  pages={4147--4153},
  year={2024},
  publisher={IEEE}
}

@article{kaltsas2022neural,
  title={Neural network tokamak equilibria with incompressible flows},
  author={Kaltsas, DA and Throumoulopoulos, GN},
  journal={Physics of Plasmas},
  volume={29},
  number={2},
  year={2022},
  publisher={AIP Publishing}
}

@article{wang2022review,
  title={A review on extreme learning machine},
  author={Wang, Jian and Lu, Siyuan and Wang, Shui-Hua and Zhang, Yu-Dong},
  journal={Multimedia Tools and Applications},
  volume={81},
  number={29},
  pages={41611--41660},
  year={2022},
  publisher={Springer}
}

@article{fiorenza2026preliminary,
  title={{Preliminary validation of ITER magnetic diagnostic algorithms by using JT-60SA magnetic measurements}},
  author={Fiorenza, F and De Tommasi, G and Frattolillo, D and Ingesson, C and Inoue, S and Kojima, S and Mattei, M and Mele, A and Miyata, Y and Neto, A and others},
  journal={Fusion Engineering and Design},
  volume={222},
  pages={115486},
  year={2026},
  publisher={Elsevier}
}

@article{sartori2003jet,
  title={{JET real-time object-oriented code for plasma boundary reconstruction}},
  author={Sartori, Filippo and Cenedese, Angelo and Milani, F},
  journal={Fusion Engineering and Design},
  volume={66},
  pages={735--739},
  year={2003},
}

@article{nishizawa2024equilibrium,
  title={{Equilibrium reconstruction of axisymmetric plasmas by combining Gaussian process regression and Markov chain Monte Carlo sampling}},
  author={Nishizawa, Takashi and others},
  journal={Plasma Physics and Controlled Fusion},
  volume={67},
  number={1},
  pages={015006},
  year={2024},
}

@article{lukovsevivcius2009reservoir,
  title={Reservoir computing approaches to recurrent neural network training},
  author={Luko{\v{s}}evi{\v{c}}ius, Mantas and Jaeger, Herbert},
  journal={Computer Science Review},
  volume={3},
  number={3},
  pages={127--149},
  year={2009},
}

@article{joung2023gs,
  title={{GS-DeepNet: mastering tokamak plasma equilibria with deep neural networks and the Grad--Shafranov equation}},
  author={Joung, Semin and others},
  journal={Scientific Reports},
  volume={13},
  number={1},
  pages={15799},
  year={2023},
}

@article{albanese1998linearized,
  title={{The linearized CREATE-L plasma response model for the control of current, position and shape in tokamaks}},
  author = {R. Albanese and others},
  journal={Nuclear Fusion},
  volume={38},
  number={5},
  pages={723},
  year={1998},
  publisher={IOP Publishing}
}

@article{albanese2015create,
  title={{CREATE-NL+: A robust control-oriented free boundary dynamic plasma equilibrium solver}},
  author = {R. Albanese and others},
  journal={Fusion Engineering and Design},
  volume={96},
  pages={664--667},
  year={2015},
  publisher={Elsevier}
}

@book{freidberg2008plasma,
  title={Plasma physics and fusion energy},
  author={Freidberg, Jeffrey P},
  year={2008},
  publisher={Cambridge University Press}
}

@book{wesson2011tokamaks,
  title={Tokamaks},
  author={Wesson, John and Campbell, David J},
  edition={Fourth},
  year={2011},
  publisher={Oxford University Press}
}

@BOOK{AriolaPironti:Springer,
  AUTHOR =       {M.~Ariola and A.~Pironti},
  TITLE =        {{Magnetic Control of Tokamak Plasmas}},
  PUBLISHER =    {Springer},
  YEAR =         {2016},
  edition = {$2^\mathrm{nd}$},
}

@article{marrelli2019upgrades,
  title={{Upgrades of the RFX-mod reversed field pinch and expected scenario improvements}},
  author={Marrelli, Lionello and others},
  journal={Nuclear Fusion},
  volume={59},
  number={7},
  pages={076027},
  year={2019},
}

@article{walker2009feedback,
  title={On feedback stabilization of the tokamak plasma vertical instability},
  author={{M.~L.}~Walker and {D.~A.}~Humphreys},
  journal={Automatica},
  volume={45},
  number={3},
  pages={665--674},
  year={2009},
}

@article{inoue2021development,
  title={{Development of JT-60SA equilibrium controller with an advanced ISO-FLUX control scheme in the presence of large eddy currents and voltage saturation of power supplies}},
  author={Inoue, S and Miyata, Y and Urano, H and Suzuki, T},
  journal={Nuclear Fusion},
  volume={61},
  number={9},
  pages={096009},
  year={2021},
}

@article{de2024control,
  title={Control of elongated plasmas in superconductive tokamaks in the absence of in-vessel coils},
  journal={Nuclear Fusion},
  author={G.~{De Tommasi} and others},
  volume={64},
  number={7},
  pages={076005},
  year={2024},
}

@techreport{ferron1997real,
  title={{Real time equilibrium reconstruction for control of the discharge in the DIII-D tokamak}},
  author={Ferron, JR and others},
  year={1997},
  institution={General Atomics, San Diego, CA (United States)},
  note= {https://www.osti.gov/servlets/purl/505400, accessed on 12.11.2025},
}

@article{huang2018improvement,
  title={{Improvement of GPU parallel real-time equilibrium reconstruction for plasma control}},
  author={Huang, Y and Xiao, BJ and Luo, ZP and Yuan, QP},
  journal={Fusion Engineering and Design},
  volume={128},
  pages={82--85},
  year={2018},
}

@article{moret2015tokamak,
  title={{Tokamak equilibrium reconstruction code LIUQE and its real time implementation}},
  author={Moret, {J.-M.} and others},
  journal={Fusion Engineering and Design},
  volume={91},
  pages={1--15},
  year={2015},
}

@article{yuan2013plasma,
  title={{Plasma current, position and shape feedback control on EAST}},
  author={Yuan, {Q.~P.} and others},
  journal={Nuclear Fusion},
  volume={53},
  number={4},
  pages={043009},
  year={2013},
}

@article{ambrosino2009design,
  title={{Design of the plasma position and shape control in the ITER tokamak using in-vessel coils}},
  author={Ambrosino, Giuseppe and Ariola, Marco and De Tommasi, Gianmaria and Pironti, Alfredo and Portone, Alfredo},
  journal={IEEE Transactions on Plasma Science},
  volume={37},
  number={7},
  pages={1324--1331},
  year={2009},
}

@article{frattolillo2025implementation,
  title={{Implementation of an ITER-relevant QP-based current limit avoidance algorithm in the TCV tokamak}},
  author={Frattolillo, Domenico and others},
  journal={Plasma Physics and Controlled Fusion},
  volume={67},
  number={5},
  pages={055017},
  year={2025},
}

@article{beghi2005advances,
  title={Advances in real-time plasma boundary reconstruction: from gaps to snakes},
  author={Beghi, Alessandro and Cenedese, Angelo},
  journal={IEEE Control Systems Magazine},
  volume={25},
  number={5},
  pages={44--64},
  year={2005},
}

@article{ambrosino2019magnetic,
  title={{Magnetic configurations and electromagnetic analysis of the Italian DTT device}},
  author={Ambrosino, Roberto and others},
  journal={Fusion Engineering and Design},
  volume={146},
  pages={1246--1253},
  year={2019},
}

@article{ariola2002design,
  title={Design and experimental testing of a robust multivariable controller on a tokamak},
  author={Ariola, Marco and Ambrosino, Giuseppe and Pironti, Alfredo and Lister, Jonathan B and Vyas, Parag},
  journal={IEEE Transactions on Control Systems Technology},
  volume={10},
  number={5},
  pages={646--653},
  year={2002},
}

@article{degrave2022magnetic,
  title={Magnetic control of tokamak plasmas through deep reinforcement learning},
  author={Degrave, Jonas and others},
  journal={Nature},
  volume={602},
  number={7897},
  pages={414--419},
  year={2022},
}

@article{dubbioso2023deep,
  title={A deep reinforcement learning approach for vertical stabilization of tokamak plasmas},
  author={Dubbioso, S and De Tommasi, G and Mele, Adriano and Tartaglione, G and Ariola, M and Pironti, A},
  journal={Fusion Engineering and Design},
  volume={194},
  pages={113725},
  year={2023},
}

@article{ambrosino2008design,
  title={{Design and implementation of an output regulation controller for the JET tokamak}},
  author={Ambrosino, Giuseppe and Ariola, Marco and Pironti, Alfredo and Sartori, Filippo},
  journal={IEEE Transactions on Control Systems Technology},
  volume={16},
  number={6},
  pages={1101--1111},
  year={2008},
}

@article{ambrosino2008plasma,
  title={{Plasma strike-point sweeping on JET tokamak with the eXtreme Shape Controller}},
  author={Ambrosino, Giuseppe and others},
  journal={IEEE Transactions on Plasma Science},
  volume={36},
  number={3},
  pages={834--840},
  year={2008},
}

@article{albanese2017iter,
  title={{ITER-like vertical stabilization system for the EAST Tokamak}},
  author={Albanese, R and others},
  journal={Nuclear Fusion},
  volume={57},
  number={8},
  pages={086039},
  year={2017},
}

@inproceedings{barabaschi1993maxfea,
  title={{The MAXFEA code}},
  author={Barabaschi, P},
  booktitle={Plasma Control Technical Meeting},
  year={1993}
}

@article{kudlacek2015real,
  title={{Real time measurement of plasma macroscopic parameters on RFX-mod using a limited set of sensors}},
  author={Kudlacek, Ondrej and Zanca, Paolo and Finotti, Claudio and Marchiori, Giuseppe and Cavazzana, Roberto and Marrelli, Lionello},
  journal={Physics of Plasmas},
  volume={22},
  number={10},
  year={2015},
  publisher={AIP Publishing}
}

@article{dubbioso2024model,
  title={Model-free stabilization via Extremum Seeking using a cost neural estimator},
  author={Dubbioso, Sara and Jalalvand, Azarakhsh and Wai, Josiah and De Tommasi, Gianmaria and Kolemen, Egemen},
  journal={Expert Systems with Applications},
  volume={258},
  pages={125204},
  year={2024},
}

@article{ding2014extreme,
  title={Extreme learning machine and its applications},
  author={Ding, Shifei and Xu, Xinzheng and Nie, Ru},
  journal={Neural Computing and Applications},
  volume={25},
  number={3},
  pages={549--556},
  year={2014},
}

@inproceedings{szepesi:IAEA2025,
    author = {Szpepesi, T. and others},
    title = {Utilizing a visible camera in the first operation phase(s) of a fusion device},
    booktitle ={30th IAEA Fusion Energy Conference},
    year = {2025},
}

@article{ambrosino2002line,
  title={On-line plasma shape identification via magnetic measurements},
  author={Ambrosino, Giuseppe and Celentano, Giovanni and Garofalo, Francesco and Glielmo, L and Pironti, Alfredo},
  journal={IEEE Transactions on Magnetics},
  volume={28},
  number={2},
  pages={1601--1604},
  year={2002},
}

\end{document}